\documentclass[5p,twocolumn]{elsarticle}
\usepackage{graphicx,amssymb,amsmath,amsfonts,color,subfigure}
\usepackage{ecrc} 

\volume{00}

\firstpage{1}

\journalname{Astroparticle Physics}

\runauth{R. Aloisio, V. Berezinsky and A. Gazizov}

\jid{Astroparticle Physics}

\begin{document}

\begin{frontmatter}

\title{Transition from galactic to extragalactic cosmic rays}

\author[Arcetri,LNGS]{R. Aloisio\corref{cor1}}
\ead{aloisio@arcetri.astro.it}
\cortext[cor1]{Corresponding author}
\author[LNGS]{V. Berezinsky}
\ead{berezinsky@lngs.infn.it}
\author[LNGS,DESY]{A. Gazizov}
\ead{askhat.gazizov@desy.de}

\address[Arcetri]{INAF, Arcetri Astrophysical Observatory, I-50125 Arcetri
 (FI), Italy} 
\address[LNGS]{INFN, National Gran Sasso Laboratory, I-67010 Assergi
 (AQ), Italy} 
\address[DESY]{DESY Zeuthen, Platanenallee 6, D-157 Zeuthen, Germany} 

\begin{abstract} %
The study of the transition between galactic and extragalactic cosmic
rays can shed more light on the end of the Galactic cosmic rays
spectrum and the beginning of the extragalactic one. Three models of
transition are discussed: ankle, dip and mixed composition models. All
these models describe the transition as an intersection of a steep
galactic component with a flat extragalactic one. Severe bounds
on these models are provided by the Standard Model of Galactic Cosmic
Rays according to which the maximum acceleration energy for Iron nuclei
is of the order of $E_{\rm Fe}^{\rm max} \approx 1\times 10^{17}$~eV. In
the ankle model the transition is assumed at the ankle, a flat feature
in the all particle spectrum which observationally starts at energy 
$E_a \sim (3 - 4)\times 10^{18}$~eV. This model needs a new high energy 
galactic component with maximum energy about two orders of magnitude above that
of the Standard Model. The origin of such component is discussed. As
observations are concerned there are two signatures of the transition: 
change of energy spectra and mass composition. In all models a heavy
galactic component is changed at the transition to a lighter or proton
component. As a result the ankle model predicts a galactic Iron
component at $E < 5\times 10^{18}$~eV, while both HiRes and Auger data
show that at $(2 - 5)\times 10^{18}$~eV primaries are protons, or at
least light nuclei. In the dip model the transition occurs at the second
knee observed at energy $(4 -7)\times 10^{17}$~eV and is characterized
by a sharp change of mass composition from galactic Iron to
extragalactic protons. The ankle in this model appears automatically as
a part of the $e^+e^-$ pair-production dip. The mixed composition model
describes transition at $E \sim 3\times 10^{18}$~eV with mass
composition changing from the galactic Iron to extragalactic mixed 
composition of different nuclei. In most mixed composition models the 
spectrum is proton-dominated and it better fits HiRes than Auger
data. The latter show a steadily heavier mass composition with
increasing energy, and we discuss the models which explain it. 
\end{abstract}


\end{frontmatter}

\section{Introduction}
\label{sec:introduction}
A clear understanding of the provenance of Cosmic Rays (CR) at different
energetic regimes, namely their galactic or extragalactic origin, is
of paramount importance in unveiling the nature of the sources of these
particles, in particular at the highest energies.

The observed energy spectrum of CR has an approximately
power-law behavior for $11$ orders of magnitude in energy with several
features that can be linked with particles propagation and acceleration.
This power-law behavior is most probably indicative of a power-law
acceleration spectra, while spectral features may be assigned to
changes in the origin of particles, their propagation and acceleration.

We shall consider these features starting from the highest energies.

If primary particles are protons, their spectrum must show a steepening 
which begins at $E_{\rm GZK} \sim 50$~EeV. It is the famous Greisen 
\cite{greisen}, Zatsepin and Kuzmin \cite{ZK} cutoff, referred to as GZK,
caused by pion photo-production energy loss in collisions of protons with 
the Cosmic Microwave Background (CMB) photons. Most probably this 
cutoff is already observed in the HiRes \emph{(High Resolution Fly's Eye)}
\cite{HiRes-GZK} and TA \emph{(Telescope Array)} \cite{TA-GZK} data,
while the spectrum steepening observed by the PAO \emph{(Pierre Auger
Observatory)} \cite{Auger-GZK} does not agree well, according 
to our calculations, with the predicted GZK shape and position. 

An alternative to the protons as primaries is given by nuclei.
Propagating through astrophysical backgrounds nuclei lose 
their energy in photo-disintegration and pair production processes, and 
also due to adiabatic expansion of the universe. The steepening of nuclei 
spectra due to photo-disintegration in the interactions with the CMB has 
been shortly mentioned by Greisen \cite{greisen} and later considered 
quantitatively in \cite{stecker69,BZ71,BGZ75,hillas75,stecker75}. The 
beginning of the steepening in this case (nuclei in the CMB radiation only) 
corresponds to the intersection of $e^-e^+$ and adiabatic energy losses 
 curves \cite{BZ71,BGZ75}. The beginning of this steepening is lower than in 
the case of GZK cutoff and the shape of this feature differs from that 
of GZK. Recent accurate calculations show that the PAO spectrum 
steepening may be explained not only by nuclei interactions. It may be 
also produced as a result of a decreasing acceleration efficiency in
the vicinity of the maximum energy that sources can provide
($E_{\max}$) (for reviews see \cite{GZK-rev}). The HiRes and 
TA data, in favor of proton composition, and the PAO data, in favor 
of nuclei composition, are further discussed in sections~\ref{sec:GZK} 
and~\ref{sec:auger}.
 
There is another feature in the spectrum of UHE extragalactic protons, 
the {\em pair-production dip} \cite{BG88}, which, as in case of GZK cutoff, 
can be directly linked to the interaction of CRs with the CMB. This dip
arises due to electron-positron production energy loss by extragalactic
protons interacting with the CMB photons: 
\begin{equation} p+\gamma_{\rm CMB} \rightarrow p+e^++e^- . 
\label{eq:pairprod} 
\end{equation} 

This feature has been studied in \cite{Stanev2000,BGGPL,BGGprd,Aletal}. 
The dip was observed with a very good statistical significance, 
$\chi^2/\mbox{d.o.f.} \sim 1$, by the Fly's Eye, Yakutsk, Akeno-AGASA 
and HiRes detectors \cite{data}, but it is absent in the Auger  
data (see section \ref{sec:auger} for a more detailed analysis).

The pair-production dip and GZK cutoff are signatures of protons. A
confirmation of the shape of these features can be considered as an
indirect evidence for a proton-dominated composition of primary CRs. 
For nuclei as primaries the shapes of the dip and cutoff are strongly
different. 

A different explanation of the dip has been proposed by Hill and Schramm
\cite{HS85}. They interpreted the dip observed in 1980s in terms of a
two-component model; the low energy component was either galactic or
produced by Local Supercluster. A similar model was later considered in
\cite{YT}. The Hill-Schramm's interpretation is widely adopted now. 

From 1960s there was observed a flattening in the Ultra High Energy 
Cosmic Rays (UHECR) spectrum, which was called the {\em ankle}. 
Discovery of this feature at the Volcano Ranch detector \cite{linsley} was 
interpreted as the transition from a steep galactic component to a flatter
extragalactic one. The ankle is clearly seen in the PAO data at
$E_a^{\rm obs} \approx 4 $~EeV \cite{auger-ankle}. In \cite{stanev2005} -
\cite{waxman} the ankle is considered as a feature where the transition
from the galactic to extragalactic cosmic rays occurs. In the dip model
\cite{BGGPL,BGGprd,Aletal} the ankle appears automatically as a part of
the pair-production dip. 

Below the ankle, at $E_{\rm skn} \approx (0.4 - 0.7)$~EeV, there
is a faint feature in the spectrum \cite{second-knee} called the {\em
second knee}. It is seen in many experiments (for a review see
\cite{2knee-ankle-rev}). This feature is also often interpreted as the
place of transition from galactic to extragalactic CRs, with the dip
model \cite{BGGPL,BGGprd,Aletal} being among these works. 

One more well known feature, the proton knee, occurs at $E_p \sim
4$~PeV. It was discovered by the MSU (Moscow State University)
array in 1958 \cite{MSU-knee}. The knee is most probably explained by
the maximum acceleration energy for protons in the Galaxy. In the
framework of the rigidity-dependent acceleration model it was predicted
that the Iron knee is located at $Z_{\rm Fe}=26$ times higher energy,
$E_{\rm Fe} \simeq 0.1$ EeV. This feature was confirmed by the KASCADE 
Grande \cite{kascadeFe}, so that the end of the Galactic CR spectrum is 
naturally connected with the Iron knee.

Based on the observational features described above, three models of the
transition from galactic to extragalactic cosmic rays have been 
proposed.

In the {\em ankle models} \cite{stanev2005} - \cite{waxman} it is assumed
that the transition occurs at the flat part of the observed spectrum
in the energy interval $E_a^{\rm trans} \sim (3 - 10)$~EeV. The
transition energy  is given by the
intersection of a flat extragalactic spectrum and a very steep galactic
one. In the majority of ankle models the extragalactic component is 
assumed to be pure proton, while the galactic one should be naturally 
represented by Iron nuclei at energies above the Iron knee. These 
models predict a transition from an Iron-dominated composition to a 
proton-dominated one at the ankle energy.

In the {\em mixed composition model} \cite{mixed} it is assumed that 
the extragalactic component consists of nuclei of various types. Thus
transition here occurs from Iron to lighter nuclei of mixed composition;
it can occur at the ankle or nearby it.

In the {\em dip model} the transition begins at the second knee and is
completed at the beginning of the dip, at $E \approx 1 $~EeV. The ankle
in this model appears as an intrinsic part of the dip. Like in the ankle
model, the transition here also occurs as an intersection of the flat
extragalactic component (this flatness is especially prominent in the
case of diffusive propagation) with a steep Galactic spectrum. In
contrast to the ankle and mixed composition models, the dip model
predicts an almost pure proton composition above $E \approx 1$~EeV and a
pure Iron composition below this energy.

In this review we shall discuss recent modifications of these three 
basic models. The paper is organized as follows: in section
\ref{endGCR} we briefly discuss the standard model of Galactic cosmic
rays, focusing in particular on the end of the spectrum, in section
\ref{UHECR} we discuss the spectra of cosmic rays at ultra high energy,
focusing on their extragalactic origin, in section \ref{transition} we
analyze the models of the transition from Galactic to extragalactic
cosmic rays. Finally, in section \ref{conclusions} we shall present our
conclusions. 

\section{The end of the Galactic CRs}
\label{endGCR}

For a long time it was assumed that Supernova (SN) explosions may
provide the observed cosmic rays. Baade and Zwicky \cite{BaaZwi} in
1934 were the first who understood that SNe energy release can produce
enough energy to fill the Milky Way with the observed cosmic rays.
Arguing with the adherents of extragalactic origin of the observed 
cosmic rays, Ginzburg and Syrovatsky put forward a model \cite{GiSy} of
Galactic Cosmic Rays (GCR). Its basic elements were particle
acceleration by SN shocks followed by a diffusive propagation in the
Galactic magnetic fields, until they leave the Galaxy.

\subsection{Standard Model for Galactic Cosmic Rays (SM GCR)}
\label{SM}
In spite of some small disturbing contradictions that will be discussed
later, one may claim now that the Standard Model for Galactic Cosmic
Rays exists (see reviews \cite{book,galprop,blasi-rev}). It is based on
the Supernova Remnant (SNR) paradigm and includes four basic elements: 
\begin{enumerate}
\item supernova remnants (SNRs) as sources;
\item SNR shock acceleration;
\item rigidity-dependent injection as a mechanism providing the
 observed CR mass composition, and 
\item diffusive propagation of CRs in the Galactic magnetic fields. 
\end{enumerate}

Let us discuss these elements in more detail.

\medskip
1. SNRs are able to provide the observed CR energy production in 
the Galaxy, which can be found \cite{book} as
\begin{equation}
Q_{\rm cr} \approx \omega_{\rm cr} c M_g/x_{\rm cr},
\label{Qcr}
\end{equation}
where $\omega_{\rm cr} \approx 0.5$~eV/cm$^3$ is the observed CR energy
density, $c$ is the velocity of CR particle, $M_g \approx 5\times 10^{42}$ g
is the total mass of galactic gas, and $x_{\rm cr} \approx 7$ g/cm$^2$
is the grammage traversed by CR before escaping the Galaxy. Using these
numbers one obtains $Q \approx 2\times 10^{40}$~erg/s, which is less
than $10\%$ of the energy release in the form of kinetic energy of the
SN ejecta per unit time. 

\medskip 
2. The most efficient mechanism of diffusive shock
acceleration (DSA) in SNRs was discovered in 1977 - 1978 \cite{DSA}. A
great progress has been reached during the last decade in the theory of
acceleration. It was shown that cosmic ray streaming instability
strongly amplifies magnetic fields upstream. It creates a highly
turbulent field with strength up to $\delta B \sim B \sim 10^{-4}$~G
\cite{Bell} (for recent works see \cite{Blasi}). At each moment of the
shock propagation only particles accelerated up to maximum energy
$E_{\rm max}$ can escape the acceleration region \cite{PZ06}. As a
byproduct, this solves the problem of adiabatic energy losses, since
particles escaping with $E_{\rm max}$ do not dwell long in the expanding
SN shell. 

$E_{\rm max}$ reaches the highest value at the beginning of the Sedov
phase and then diminishes due to the shock deceleration. At each moment
$t$ the spectrum of escaping particles has a narrow peak at $E_{\rm
max}(t)$, but the spectrum integrated over time acquires a classical
$E^{-2}$ shape with a flattening at highest energies \cite{PZ06}. The
maximum acceleration energy estimated in the Bohm regime of diffusion is
given by
\begin{equation}
E_{\rm max} = 4 Z \frac{B}{10^{-4}{\rm G}}
\left ( \frac{W_{51}}{n_g/{\rm cm}^{3}} \right )^{2/5}{\rm PeV},
\label{eq:Emax}
\end{equation} 
where $B$ is the amplified magnetic field, $W_{51}$ is the kinetic
energy of the shell in units $10^{51}$~erg, $n_g$ is the upstream
density of the gas and $Z$ is the charge number of accelerated nuclei.
Thus for protons and Iron nuclei maximum energies are
\begin{equation}
\begin{array}{ll}
\displaystyle E_p^{\rm max} &= ~~4\times B_{-4} {\rm ~PeV,}\\
\displaystyle E_{\rm Fe}^{\rm max}&\simeq 0.1\times B_{-4} {\rm~ EeV.}
\end{array} 
\label{eq:xdef}
\end{equation}
Here $E_p^{\rm max}$ describes well the position of the proton knee and 
$E_{\rm Fe}^{\rm max}$ predicts the position of the Iron one.

An important result on the structure of the knee was obtained
in the KASCADE-Grande experiment \cite{kascadeFe}. The spectra were
measured separately for electron-rich showers, initiated by protons and
light nuclei, and electron-poor showers, corresponding to the heaviest
nuclei of the Iron group. The steepening in the spectrum of 
electron-poor showers was found at $E \approx 80$~PeV. With
electron-rich showers knee at $(3 - 5)$~PeV, the heaviest nuclei knee
corresponds to Iron nuclei, as expected in the rigidity-dependent
acceleration model. 

\medskip 
3. Observations show that nuclei in cosmic rays are
systematically more abundant than in the interstellar medium of the
solar neighborhood \cite{Ellison,Meyer,BK99}; this occurs due to the
injection of particles in the regime of acceleration
\cite{Ellison,Meyer,BK99}. It can be illustrated by the following simple
consideration \cite{Aletal}. A particle $i$ from downstream ($i=A,~p$)
can cross the shock and thus may be injected in the regime of
acceleration if its Larmor radius $r_L(p) \geq d$, where $d$ is the
thickness of the shock front. It readily gives the relation
between a nucleus and proton injection momenta
\begin{equation} 
p_{\rm inj}^A = Z e B d /c = Z p_{\rm inj}^p.
\label{eq:p_inj}
\end{equation}
Therefore, $v_{\rm inj}^A < v_{\rm inj}^p$, which provides a higher
injection rate for nuclei. 

The same result may be also obtained in a more formal way. Consider a 
flux of accelerated particles $i$, $J_i(p)=K_i (p/p_{\rm inj}^i)^{-\gamma_g}$.
Normalizing $J_i(p)$ by the condition 
\begin{equation}
\frac{4\pi}{c}\int_{p_{\rm inj}^i}^\infty J_i(p) dp = \eta_i n_i, 
\label{eq:norm}
\end{equation}
where $n_i$ is the density of gas $i$ and $\eta_i$ is the fraction of 
this density injected into acceleration process, one obtains a
ratio of nuclei and proton CR fluxes 
\begin{equation}
\frac{J_A(p)}{J_p(p)}=Z^{\gamma_g-1}\frac{\eta_A n_A}{\eta_p n_p}.
\label{eq:fraction}
\end{equation}
Thus, the fraction of nuclei is enhanced by a factor $Z \eta_A/\eta_p$. 
For numerical calculations of CR nuclei abundances see 
\cite{Ellison,Meyer,BK99}.

\medskip
4. CRs propagate in the Galaxy diffusively; they scatter off small-scale
magnetic turbulence which may be described as a superposition of MHD
waves with different amplitudes and random phases. This process is
considered (see e.g.\ \cite{book}) in the resonance approximation, when
the particle giro-frequency is equal to the wave frequency in the system
at rest with the particle motion along the average magnetic field. The
magnetic field may be considered as a combination of an average
(constant) field, $\vec{B}_0$, and a fluctuating component, $\vec{B}$.
In \cite{book} the parallel diffusion coefficient $D_{\parallel}(E)$ is
calculated assuming $D_{\perp}(E) \ll D_{\parallel}(E)$ (see however 
numerical simulations \cite{DBS07} which do not support this assumption 
at the highest energies).

\begin{figure*}[t]
\begin{center}
\includegraphics [width=0.515\textwidth]{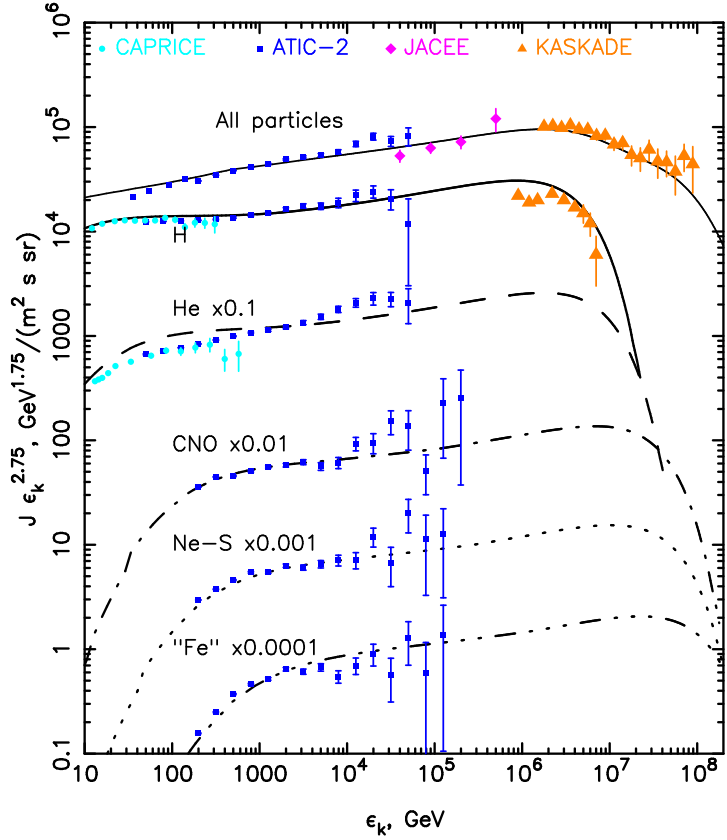}
\end{center}
\caption{Fluxes and spectra calculated within the Standard Model
\cite{BV07} for all particles, protons and nuclei. Fluxes are shown as
function of kinetic energy $\epsilon_k$ per nucleon. They are compared
with the data of CAPRICE, ATIC-2, JACEE and KASCADE. The position of
the knees for all nuclei are given by $\epsilon_{\rm kn} \approx 3
Z$~PeV. The end of Galactic spectrum is given by the iron knee
$\epsilon_k \approx 80$~PeV. At higher energies Galactic spectrum
becomes very steep.}
\label{fig:b-volk}
\end{figure*}

Proton and nuclei spectra, as calculated by Berezhko and V\"olk
\cite{BV07} within SM GCR, are shown in Fig.~\ref{fig:b-volk} in
comparison with data. An agreement with observations is quite good at
low energies, and the knee is confirmed at $E_{\rm kn} \approx 3$~PeV in
proton and all-particle spectra. The Iron knee located at $E_{\rm Fe}
\approx 80$~PeV is the most important prediction of the Standard Model;
this feature was confirmed by the KASCADE-Grande detector
\cite{kascadeFe}. The spectra beyond knees are expected to be very
steep.

\subsection{Diffusive propagation: qualitative estimates}
\label{sec:diff-estimates}

Diffusive propagation is a phenomenon which currently provides problems
to the Standard Model of Galactic CRs. The essence of these problems can
be easily seen (for a detailed discussion see also the papers of other
authors that cover these topics in the present topical issue).

The diffusion coefficient and its energy dependence are primarily 
determined by the turbulence spectrum $w(k)$, which in the most
important cases is given in a power-law form $w(k) \propto k^{-m}$,
where $k$ is the wave number. Then one has 
\begin{equation} 
D(E) \propto E^{\mu},\;\; \mu = 2 - m,
\label{eq:D(E)}
\end{equation}
with $\mu=1/3$,~ $\mu=1/2$ and $\mu=0$ for the Kolmogorov, Kraichnan
and shock wave diffusion, respectively. In the Bohm regime with 
strong turbulence, the diffusion length is of the order of the gyroradius 
and $\mu=1$.

It is easy to verify that the space density $n_p$ of primary protons and
nuclei is 
\begin{equation}
n_p(E) \propto E^{-(\gamma_g+\mu)}. 
\label{n_p}
\end{equation} 
It follows from $n_p(E) \sim Q_p(E) \tau_{\rm esc}(E)$, where $Q_p(E)
\propto E^{-\gamma_g}$ is the generation (acceleration) rate of
primaries, i.e.\ protons and nuclei, and $\tau_{\rm esc}(E) \propto
D(E)^{-1}$ is the escape time from the Galaxy. For non-relativistic
diffusive acceleration $\gamma_g=2$; hence from the index of the
observed spectrum, $\gamma_g +\mu =2.7$, one derives $\mu=0.7$. At very
high energy this results in a too large anisotropy, $\delta(E) \propto
D(E) \propto E^{\mu}$, and in a too small traversed grammage, $X_{\rm
cr}(E) \propto 1/D(E)$. 

To solve the problem of small $X_{\rm cr}(E)$ it was suggested the
effect of spallation inside CR sources and re-acceleration by second
order Fermi mechanism during CR propagation \cite{PtuskinDrury}. 

Concerning anisotropy, observations show its constant behavior with
energy which is in contrast with the simple estimate above. This
inconsistency is only apparent because the result $\delta(E)\propto
D(E)$ refers to an average of the anisotropy amplitude computed over
many source realizations, i.e. over a continuum of sources. In the
framework of the SM for GCR, the actual value of anisotropy can be
computed only taking into account the random nature (in space and time)
of nearby recent SNRs \cite{BlasiAmato}.

It is easy to estimate the $e^+/e^-$ ratio in the framework of SM and
to compare it with the one measured in the PAMELA experiment
\cite{pamela-positrons}.

The basic difference with protons and nuclei, considered above, is that
the lifetime of HE electrons and positrons is much shorter than the
escape time. It is given by energy losses due to synchrotron and
inverse Compton radiation, $\tau_e \propto E^{-1}$. For primary HE
electrons the rate of production (acceleration) is $Q_e(E) \propto
E^{-\gamma_g}$ and the space density 
\begin{equation}
n_e(E) \sim Q_e(E) \tau_e(E) \propto E^{-(\gamma_g+1)}.
\label{n_e}
\end{equation}
Positrons in the SM GCR are produced as secondary particles in $p+p \to
\pi^+ \to e^+$ interactions of CR protons with the interstellar gas.
Using $n_p(E)$ from Eq.(\ref{n_p}), one finds the rate of positron
generation as
\begin{equation}
Q_{e^+}(E) \sim n_p(E) \sigma_{pp} n_{\rm gas} c \propto 
E^{-(\gamma_g+\mu)}. 
\label{Qe+}
\end{equation}
The space density of positrons $n_{e^+}(E)$ is then obtained as 
\begin{equation}
n_{e^+}(E) \sim Q_{e^+}(E) \tau_e \propto E^{-(\gamma_g +\mu+1)}.
\label{ne+}
\end{equation}
Thus in the framework of the SM of GCR the ratio 
\begin{equation}
n_{e^+}/n_{e-} \propto E^{-\mu}
\label{ratio}
\end{equation}
regularly diminishes with energy increasing. This simple estimate is in
contrast with the observed increasing of the $e^{+}/e^{-}$ ratio in the
PAMELA data \cite{pamela-positrons}. 

\subsection{Standard Model: Problems and Solutions}
\label{sec:ProblemsSolutions} %
The SM of Galactic cosmic rays faces at least two major problems. The
first one is connected with the diffusion mechanism and was already
discussed in the previous section \ref{sec:diff-estimates}. The second
one is connected with the acceleration mechanism of CRs and will be 
discussed below. Here we shall address only the contradictions of the 
SM of GCR with observations, demonstrating that in most cases they are
only apparent. Numerous subdominant phenomena, like re-acceleration,
which in fact may affect the results in a noticeable way, will be
omitted.

As far as {\em diffusion} is concerned, the problems arise due to the
fast increase of the diffusion coefficient with energy $D(E) \propto
E^{\mu}$, which results at high energy in a large anisotropy, $\delta
(E) \propto E^{\mu}$, and in a too small traversed grammage, $X_{\rm
cr} \propto E^{-\mu}$. A possible solution was mentioned in section
\ref{sec:diff-estimates}.

It is interesting that another {\em diffusion} problem, the observation 
by PAMELA of the rise with energy in the $e^+/e^-$ ratio, may be solved
within the SM \cite{blasi-positrons} (see, however, the criticism of
this model in \cite{KOT}). The basic idea presented in
\cite{blasi-positrons} is the production of $e^+e^-$-pairs in
pp-collisions \emph{in situ}, i.e.\ at the shock front. Secondary
positrons are immediately involved in the process of acceleration, in
fact more efficiently than primary electrons, because they already have
a power-law energy spectrum. As a result, the ratio $e^+/e^-$ rises
with energy. Of course, there were proposed other mechanisms of positron
production, e.g.\ from pulsars \cite{AharAtoyVolk,MPohl}, from Dark
Matter annihilation etc.\ (see \cite{e+review} for review and
references), but it is important that the positron excess can be
successfully explained in the framework of the SM GCR.

The difference in spectrum slopes for protons and Helium was considered
for a long time as a serious problem for shock acceleration, which is
the most important component of the SM. This difference is presented in
the most precise way by the PAMELA experiment \cite{pamela-pHe}; the
spectrum of Helium is harder. The slope difference is small, but
statistically it is reliably provided, $\gamma_p - \gamma_{\rm
He}=0.1017 \pm 0.0014$.

Theoretically, the shock-wave acceleration does not distinguish among 
nuclei with equal rigidities. The predicted spectrum is a decreasing 
power-law with an exponent determined by the shock properties, i.e.\ 
the Mach number. However, according to PAMELA, spectra are not exactly 
power-law and the effective exponents are also different: $2.8$ and 
$2.7$ for protons and Helium, respectively. The understanding of these 
phenomena needs the inclusion of particle injection in the 
acceleration regime and the account of the escaping particles from 
the accelerator. We shall follow here the paper by Drury 
\cite{drury2010} with a clear physical analysis and the works 
\cite{malkov,ohira} for calculations. 

In the standard approach the spectrum is calculated for particles
accompanying the shock front, i.e.\ located within the acceleration
length upstream and downstream. This spectrum differs from that of the
escaping and captured particles. The latter drift downstream and are
trapped there until the shock dies. They suffer adiabatic energy losses
and do not contribute to the HE part of the observed spectrum. For
escaping particles the exit model must be specified. In section \ref{SM}
we considered the exit mechanism from \cite{PZ06}: at each time $t$
particles accelerated up to maximum energy $E_{\max}(t)$ freely escape
from the shell. The total spectrum of escaping particles may be
calculated by an integration over time. The highest energy particles
escape at the beginning of the Sedov phase; the lower energy ones exit
progressively later. The chemical composition changes with time because
swept-up gas changes its chemical composition and ionization state. It
means that the energy spectrum of nuclei with fixed (Z, A) is
characterized by its own $\gamma_g$.

According to section \ref{SM}, injection plays a similar role. An
account for propagation through interstellar medium can also result in a
difference in the spectral indices of protons and Helium.

\subsection{Beyond the SM: New component of the Galactic Cosmic Rays?}
\label{NewComponent} 

The observed knees are interpreted as the end of the Galactic CRs.
Positions of knees define maximum acceleration energies, different for
protons and nuclei, see Eq.~(\ref{eq:xdef}). Though this picture looks
now very natural, the way to its adoption was rather dramatic. Soon
after the discovery of DSA \cite{DSA} it was understood \cite{cesarsky}
that the maximum acceleration energy in this approach does not exceed
$Z\times 10$~TeV, much below the knee. Only later after the works by Bell 
\cite{DSA}, and Bell and Lucek \cite{Bell}, it was realized that an
increasing of the magnetic field due to the Weibel instability may
enhance the maximum energy up to the needed level 
$E_{\rm Fe}^{\max} \approx 0.1$~EeV given by Eq.~(\ref{eq:xdef}). This 
story teaches us to be careful with assumptions about higher $E_{\max}$. 
Nevertheless, reasonable mechanisms to enhance $E_{\max}$ have been 
recently found.

The motivation for an additional component of the GCRs with $E_{\max}$
up to  $10$~EeV is given by the interpretation of the ankle 
observed according to the latest data at (3 - 4)~EeV, as the
transition from Galactic to extragalactic CRs at 
$E_{\rm tr} \sim (3 - 10)$~EeV. It was proposed in 
\cite{hillas2005,hillas2006,gaisser} that such an additional component
may appear due to some unspecified sources different from SNR.

A plausible idea for increasing $E_{\max}$ is given by the explosion of
SN in the stellar wind left behind by presupernova
\cite{biermann88,BP89}. The large magnetic field in the wind and the
presence of heavy nuclei result in a larger $E_{\max}$. In particular in
\cite{biermann88} this effect is maximized by considering the Wolf-Rayet
stars as presupernovae. 

In a more general way this idea was considered in \cite{sveshnikova}. 
In this work an approximate relation between $E_{\max}$ and a SN total
energy output ${\cal E}$ is obtained as $\log E_{\max} \approx 0.52
\log {\cal E}_{51} +2.1$~TeV for the protons, where ${\cal E}_{51}$ is
the total energy output in units $10^{51}$~erg. The distribution of 
SNe over $E_{\max}$ is obtained using the distribution of SNe over 
${\cal E}_{51}$. The high energy tail of the distribution gives 
sources with enhanced $E_{\max}$, and thus the diffuse spectrum at
large $E$ extends up to $E \sim 1$~EeV for Iron nuclei. 
\begin{figure}[ht]
\begin{center}
\includegraphics [width=0.40\textwidth]{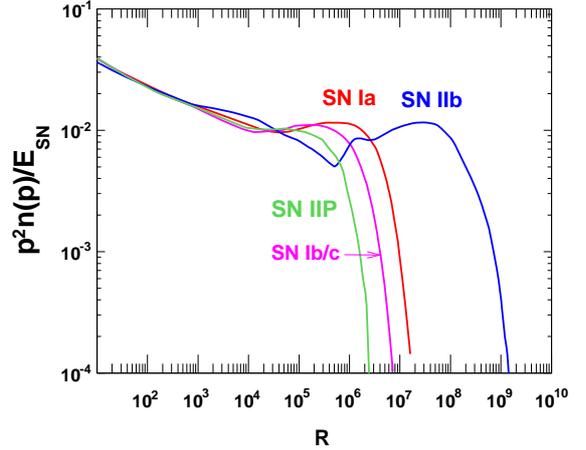}
\end{center}
\vspace{-4mm}
\caption{
Diffuse spectra of Galactic CRs produced by four groups of SNRs,
labeled as SN Ia, SN IIP, SN Ib/c and SN IIb (see \cite{PZS2010}).
Along the ordinate axis the plotted quantity is $p^2 n(p)/E_{\rm SN}$,
where $p$ is the particle momentum, $n(p)$ is proportional to the total
number of accelerated particles with momentum $p$ produced by SN and
$E_{\rm SN}$ is the total energy of SN explosion. Note that we
generalized the original plot from \cite{PZS2010} introducing
dimensionless rigidity $R=p/(Z m_N)$ instead of dimensionless proton
momentum $p/m_N$, as the authors recommended to do for nuclei. As seen
from the plot, in the maximum of the curve the rigidity reaches
$R_{\max}=3\times 10^7$, i.e.\ $0.78$~EeV for Iron (see text).
}
\label{fig:PZS}
\end{figure}

In \cite{PZS2010} a similar approach was used for the calculation 
of GCR diffuse flux from SNRs with large $E_{\max}$. SNRs were 
subdivided into four groups according to the type of explosion: 
SNR Ia, SNR IIb/c, SNR IIP and SNR IIb. Each group of SNRs was 
characterized by fixed parameters of SN explosion and presupernova 
stellar wind, in which the shock propagates, accelerating particles. 
In Fig.~\ref{fig:PZS} spectra of accelerated particles and $R_{\max}$ 
are shown for each calculated group. The group SN IIb represents 
a new HE component with a higher $E_{\max}$. This energy corresponds 
to the shock propagation through a presupernova stellar wind. For 
the evolution of SN IIb we use calculations published in 
\cite{nozawa2010}. The presupernova of SN IIb is a red supergiant 
with mass $18 M_\odot$ on the Main Sequence. The stellar wind is 
presented by Helium ($Z=2$) and at a smaller extent by Oxygen 
($Z=8$), since metals have very small abundances in presupernova 
stellar wind. The metals produced by SN explosion work as a piston, 
and the shock propagates ahead of this layer. Using Oxygen we 
obtain at the maximum of the curve $E_{\max}=Z R_{\max} \approx 
0.2$~EeV, and for Iron which has very low abundance $E_{\max} 
\approx 0.8$~EeV, still below the ankle.

The next logical step in the discussion of SN explosions as sources of a
new HE Galactic component is given by the {\em hypernova explosion}, a
very rare (one per $10^5$ years in a galaxy) but very powerful event
which can alone fill the Milky Way with cosmic rays of the observed
density. Hypernovae are assumed to be the sources of GRBs. Such a model
for GCR was considered in \cite{dermer}. The explosion occurred $2\times
10^5$~yr ago at distance about $500$ pc from the Sun. At present the
highest energy particles passed the Sun, while the lower energy
ones are still confined there. In this model the transition to
extragalactic CRs occurs at $E \sim 0.2$~EeV, too low for the ``ankle
transition''. 

An extreme case is considered in \cite{kusenko}. A sample of $1000$
hypernovae are exploding randomly in the Milky Way with time interval
$10^5$ years. They produce cosmic rays with a mass composition of 
$10\%$ protons and $90\%$ Iron nuclei with maximum energy $100$~EeV. 
Protons, propagating rectilinearly, leave the Galaxy starting from $E 
\sim 1$~EeV, while Iron nuclei diffuse in the Galactic magnetic fields 
and well explain the energy spectrum of the PAO starting from $10$~EeV. 
At lower energy the spectrum and mass composition are explained as a
mixture of protons and Iron nuclei. This is the case when hypernovae can
explain the whole spectrum of UHECR at $E > 2$~EeV by the Galactic CR
component.

To conclude, the ankle model needs a new high energy Galactic component
at $0.1{\rm ~EeV} \lesssim E \lesssim 10$~EeV with a heavy mass
composition. Though in principle it looks possible, for SNe such a case
was not found. Probably other Galactic sources, such as pulsars
\cite{book}, could fit this hope better. 

\section{Extragalactic UHECR}
\label{UHECR}

In this section the study of the transition will be proceeded moving
from the end of the UHECR spectrum (from GZK cutoff in the case of 
protons) towards lower energies. 

In the case of protons there are two spectral signatures of their
propagation through CMB: {\em pair-production dip}, which is a rather
faint feature caused by the pair-production process given by
Eq.~(\ref{eq:pairprod}), and a sharp steepening of the spectrum caused
by pion photo-production process 
\begin{equation}
p + \gamma_{\rm CMB} \rightarrow \pi + X ,
\label{eq:pgpN}
\end{equation}
called {\em GZK cutoff} \cite{greisen,ZK}. The GZK cutoff position is
roughly defined by the energy where the pair-production energy loss in
Eq.~(\ref{eq:pairprod}) and the pion production one of Eq.~(\ref{eq:pgpN})
become equal, namely at $E_{\rm GZK} \simeq 50$~EeV \cite{BG88}.

In the case of UHE nuclei the situation is different. For heavy enough
nuclei $A > 10$, taking into account only the CMB background, the
steepening of the spectrum starts at the energy where adiabatic energy
losses (due to the expansion of the universe) and losses due to pair
production become equal (see energy losses in \cite{AloNuclei}), namely
at $E \sim 20 - 50$~EeV. The pair production process experienced by nuclei 
is affected only by the CMB radiation, while the photo-disintegration
process takes place also on the Extragalactic Background Light
(EBL) \cite{SteckerEBL}. The latter is the process in which a nucleus
with atomic mass number $A$ loses one or more nucleons interacting
with the CMB or EBL:
\begin{equation} 
A+\gamma_{CMB,EBL}\to (A-nN) + nN
\label{eq:photodis}
\end{equation}
The dominant process is one nucleon emission $(n=1)$ as discussed in
\cite{AloNuclei}. 
\begin{figure}[ht]
\begin{center}
\includegraphics [width=0.40\textwidth]{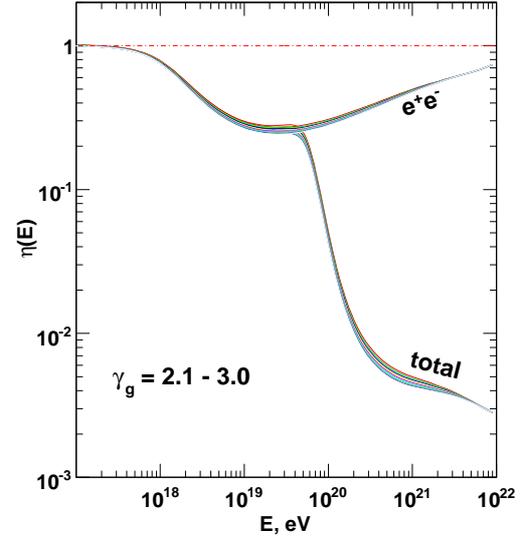}
\end{center}
\vspace{-12mm}
\caption{Modification factors for different generation indices
 $\gamma_g$.} 
\label{fig:mfactor}
\end{figure} 
The photo-disintegration of nuclei, together with the
pair production process, produces a steepening in the observed
spectrum. The exact position of the flux suppression and its shape
depend on the nuclei species as well as on the details of the
cosmological evolution of the EBL which, in contrast to CMB, is not
known analytically, being model dependent \cite{SteckerEBL}. In general
the steepening of the nuclei spectrum is not as sharp as the GZK and it
occurs at lower energies \cite{AloNuclei}.

Thus, the GZK cutoff is the most spectacular feature in the UHECR
spectrum. However, it is valid only for UHE protons and the predicted 
shape of the GZK feature is model-dependent: it depends on a possible 
local over-density or deficit of sources and can be mimicked by low 
values of the maximum energy $E_{\max}^{\rm acc}$, that sources can provide, 
In contrast, the pair-production dip
is practically model-independent mainly because protons contributing 
to this energy range arrive from distant cosmological sources.

\subsection{Pair-production dip}
\label{dip}

Being a quite faint feature, the pair-production dip is not seen well 
in the spectrum presented as $\log J(E)$ versus $\log E$. It is more
pronounced when analyzed in terms of the {\em modification factor}
\cite{BG88,Stanev2000,BGGprd,Aletal}
\begin{equation}
\eta(E)=J_p(E)/J_p^{\rm unm}(E),
\label{eq:eta}
\end{equation}
which is defined as the ratio of the spectrum $J_p(E)$, calculated with
all energy losses taken into account, and the unmodified spectrum
$J_p^{\rm unm}(E)$, where only adiabatic energy losses are included. 

\begin{figure*}[t]
\begin{center}
 \begin{minipage}[ht]{53mm}
 \includegraphics[width=53mm,height=51mm]{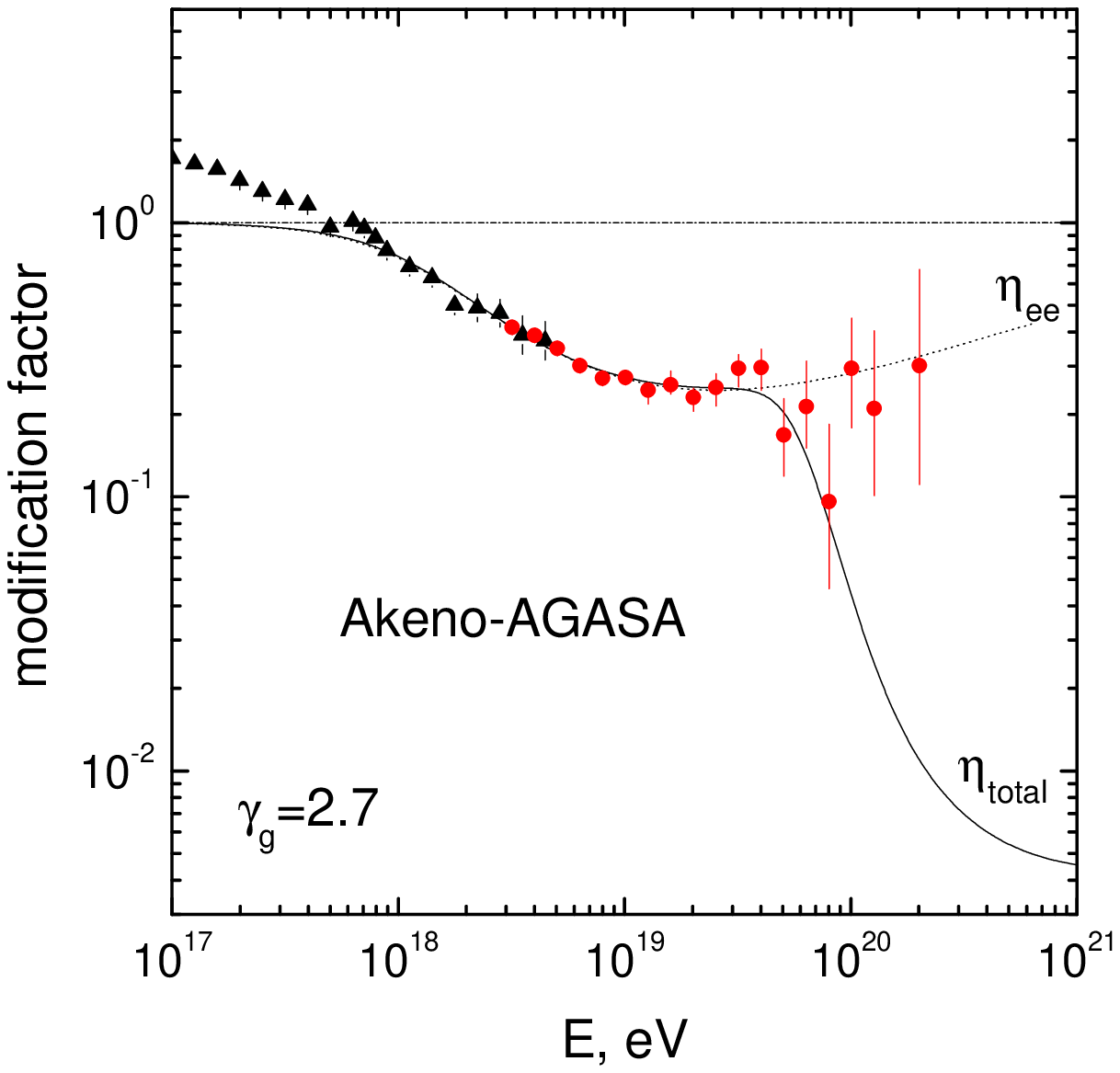}
 \end{minipage}
 \hspace{1mm}
 \begin{minipage}[h]{54 mm}
 \includegraphics[width=54mm]{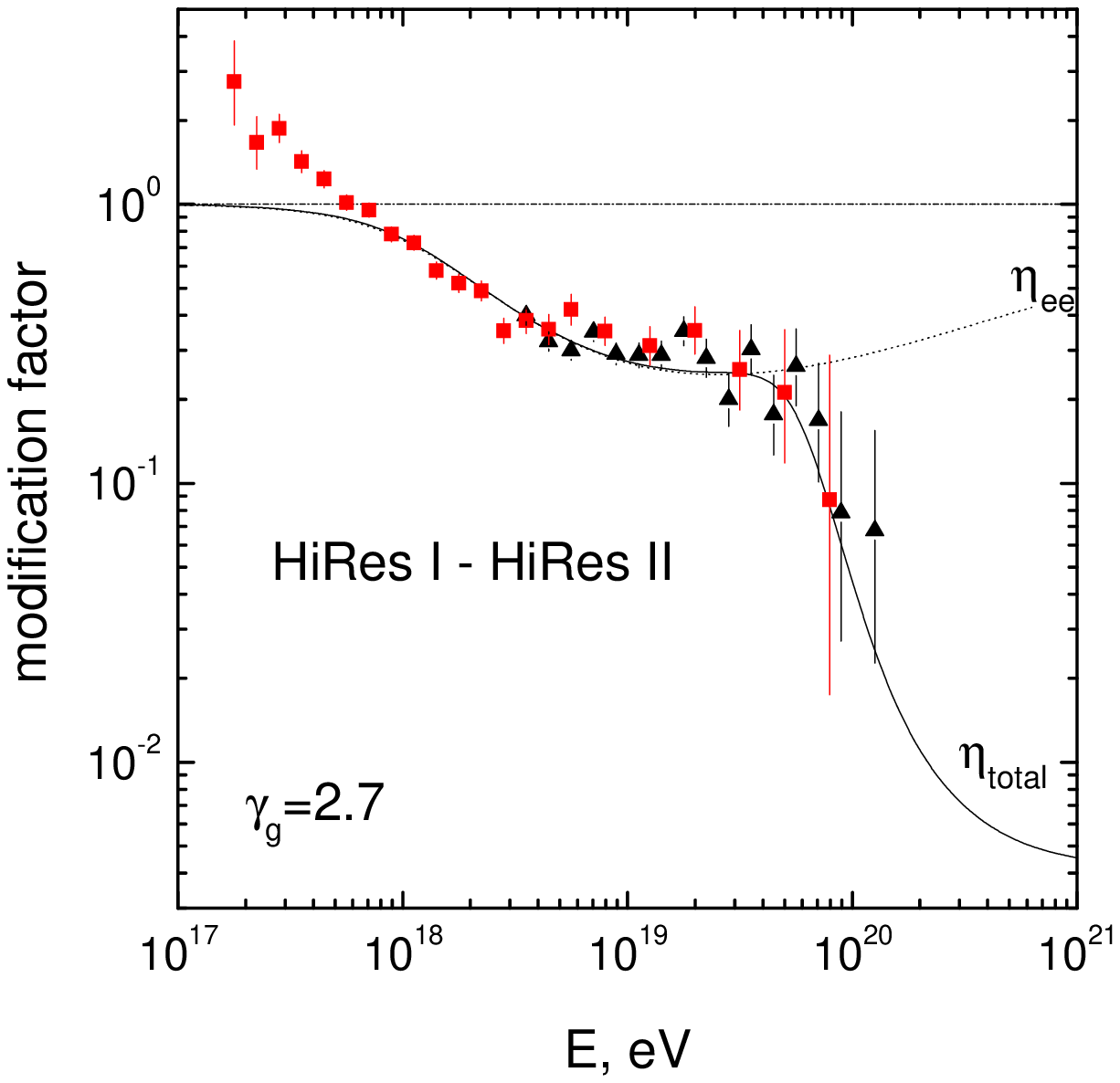}
 \end{minipage}
\newline \noindent
\medskip \hspace{-25mm}
 \begin{minipage}[ht]{54mm}
 \includegraphics[width=53mm,height=53mm]{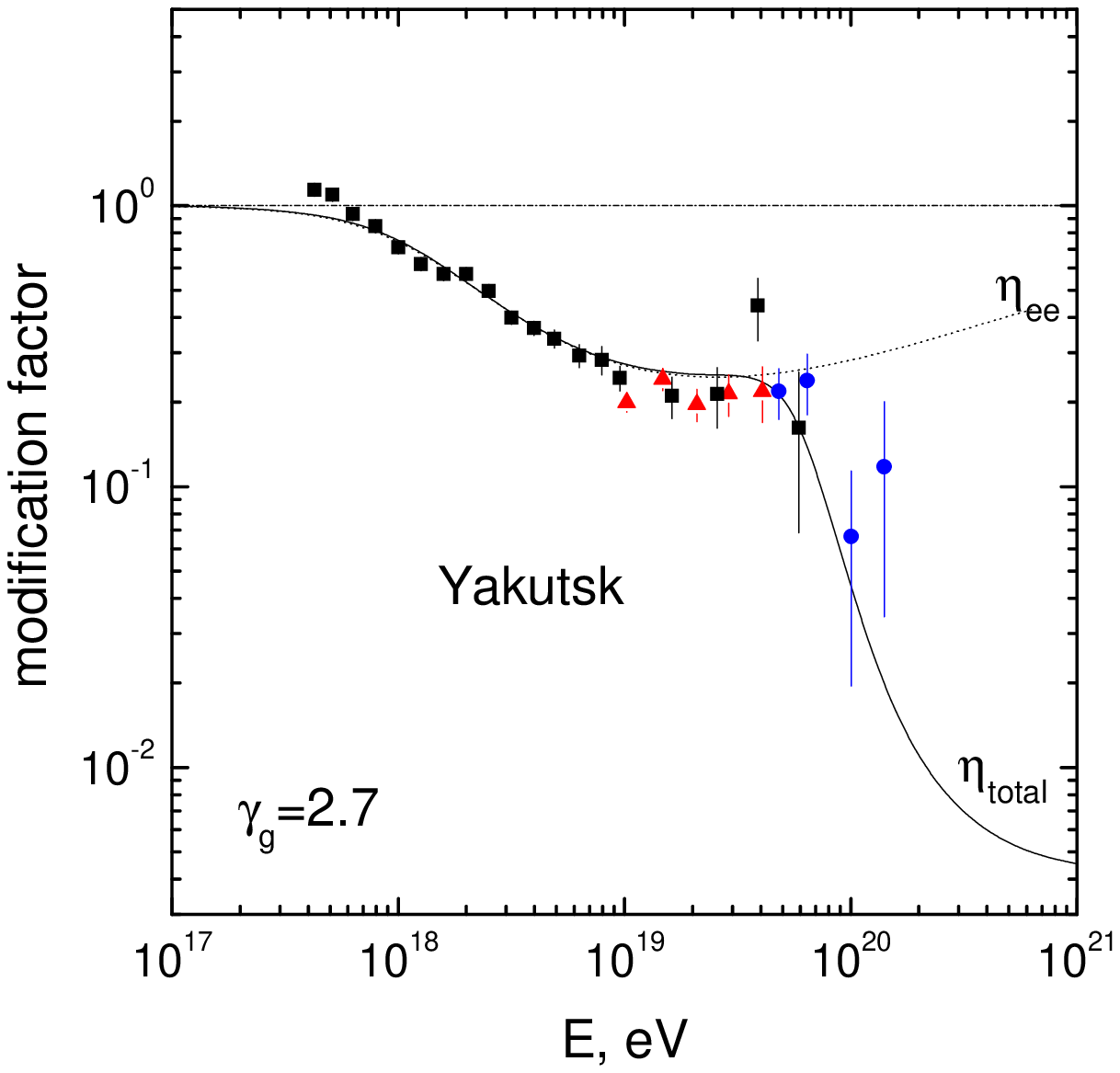}
 \end{minipage}
 \hspace{1mm}
 \begin{minipage}[h]{54 mm}\medskip
 \includegraphics[width=53mm,height=54mm]{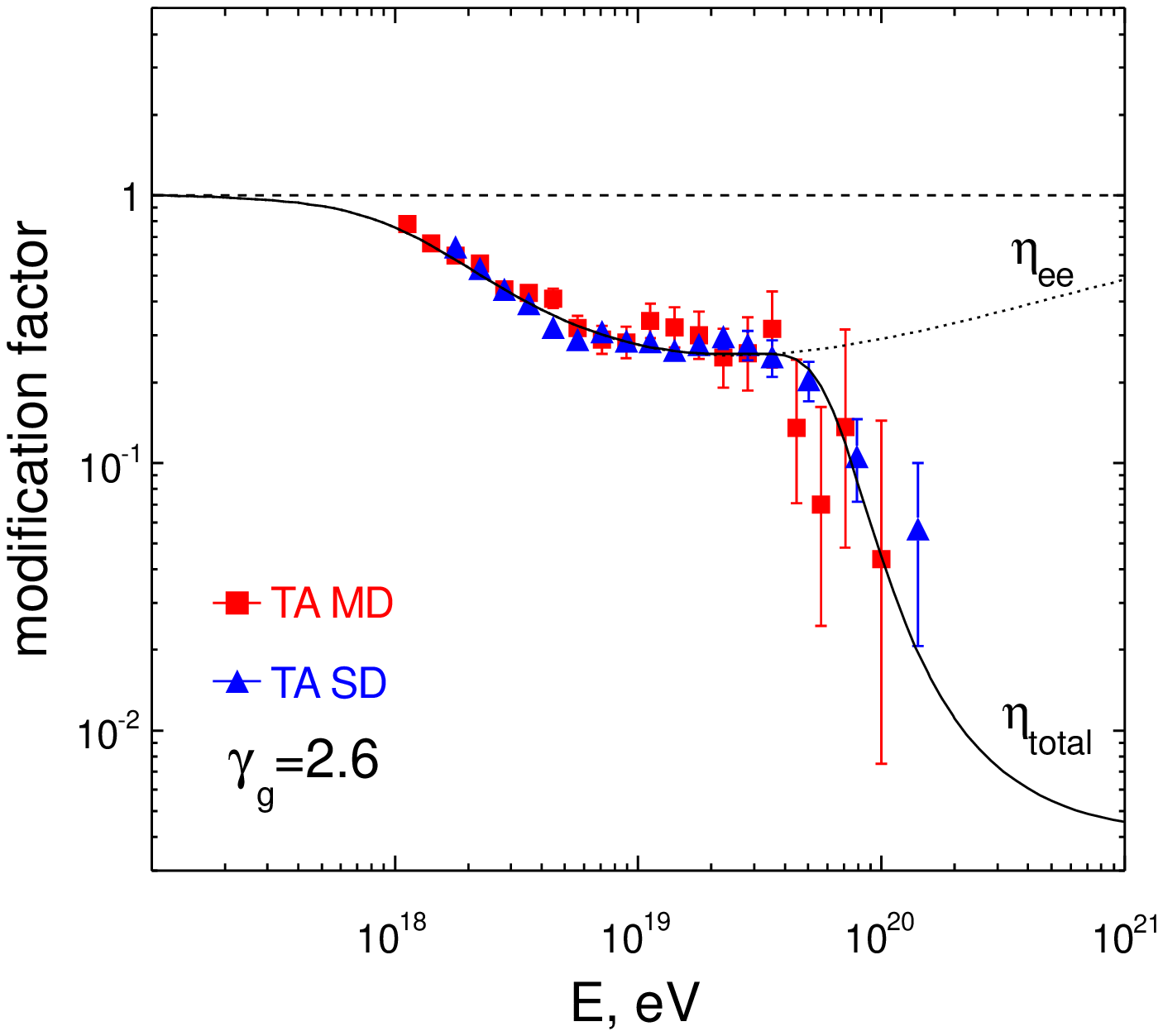}
\end{minipage}
\end{center}
\vspace{-4 mm}%
\caption{
The predicted pair-production dip in comparison with Akeno-AGASA, HiRes,
Yakutsk and Telescope Array (MD and SD) data \cite{data}. All these
experiments confirm the dip behavior with good accuracy, including also 
the data of Fly's Eye \cite{data} (not presented here).
} %
\label{fig:dips}
\end{figure*}

Theoretical modification factors calculated for different source
generation indices $\gamma_g$ are presented in
Fig.~\ref{fig:mfactor}. If one includes in the calculation of $J_p(E)$
only adiabatic energy losses, then, according to its definition, $\eta(E)=1$
(dash-dot line). When the $e^+e^-$-production energy loss is
additionally included, one obtains the $\eta(E)$ spectrum shown in
Fig.~\ref{fig:mfactor} by the curves labeled ``$e^+e^-$''. With the pion
photo-production process being also included, the GZK feature (curves
``total'') appears. The observable part of
the dip extends from the beginning of the GZK cutoff at $E \approx
40$~EeV down to $E \approx 1$~EeV, where $\eta \approx 1$. It has two
fattenings: one at energy $E_a^{\rm tr} \sim 10$~EeV and the other at $E_b
\sim 1$~EeV. The former automatically produces the ankle (see
Fig.~\ref{fig:dips}) and the latter provides an intersection of the flat
extragalactic spectrum at $E \leq 1$~EeV with the steeper Galactic
one.
Since many non-essential factors in the numerator and denominator of
Eq.~(\ref{eq:eta}) compensate or cancel each other, the dip in terms of
the modification factor is a less model dependent physical quantity than
the spectrum. In fact the dip is determined mostly by the interaction of
protons with the CMB photons and it depends mainly on the CMB spectrum 
and the differential cross-section of $e^+e^-$ pair-production. In
particular it depends weakly on the spectral index of the generation
spectrum. In Fig.~\ref{fig:mfactor} curves are plotted for $2.1 \leq
\gamma_g \leq 3.0$ with a step $\Delta\gamma_g =0.1$, and uncertainties
are seen as thickness of the curves.

Modification factors in Fig.~\ref{fig:mfactor} are presented for the
case without cosmological evolution of the sources, which is usually
described in the injection spectrum by a factor $(1+z)^m$ up to $z_{\max}$. 
The inclusion of evolution may noticeably change the modification factor, 
but in fact it allows to improve the agreement of the dip with data due to 
the additional free parameters $m$ and $z_{\max}$ (see Fig.~14 of
Ref.~\cite{BGGprd}).

Thus, a remarkable property of the dip in terms of modification factor
is its {\em universality}. The dimensionless quantity
$\eta(E)$ remains the same with various physical phenomena being
included in calculations \cite{BGGprd,Aletal}: discreteness in the
source distribution (distance between sources may vary from $1$ Mpc to
$60$ Mpc), different modes of propagation (from rectilinear to
diffusive), local overdensity or deficit of sources, large-scale
inhomogeneities in the sources distribution, some regimes of
cosmological evolution of sources (most notably those observed for AGN)
and fluctuations in the interactions. The phenomenon which modifies the
dip significantly is the possible presence of more than $15\%$ of
nuclei in the primary radiation. Therefore, the shape of the proton
dip in terms of modification factor is determined mostly by the
interaction with the CMB.

Above the theoretical modification factor was discussed. The {\em
observed} modification factor, according to definition, is given by the
ratio of the observed flux $J_{\rm obs}(E)$ and unmodified spectrum
$J_{\rm unm}(E) \propto E^{-\gamma_g}$, defined up to normalization as:
\begin{figure*}[ht]
\begin{center}
 \begin{minipage}[ht]{60 mm}
 \centering
 \includegraphics[width=59mm,height=46mm]{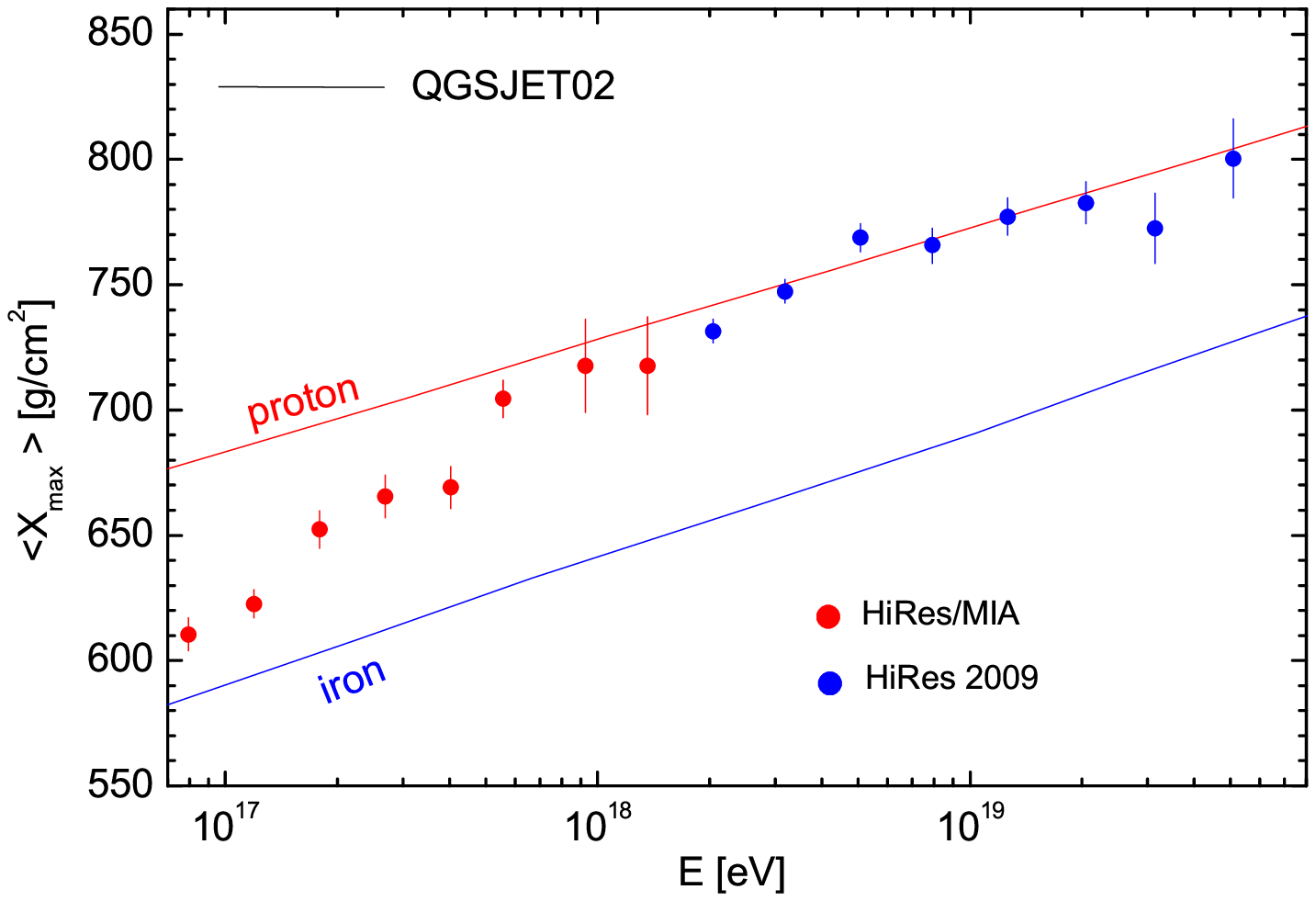}
 \end{minipage}
 \hspace{1mm}
 \vspace{-1mm}
 \begin{minipage}[h]{60 mm}
 \centering
 \includegraphics[width=60mm,height=51mm]{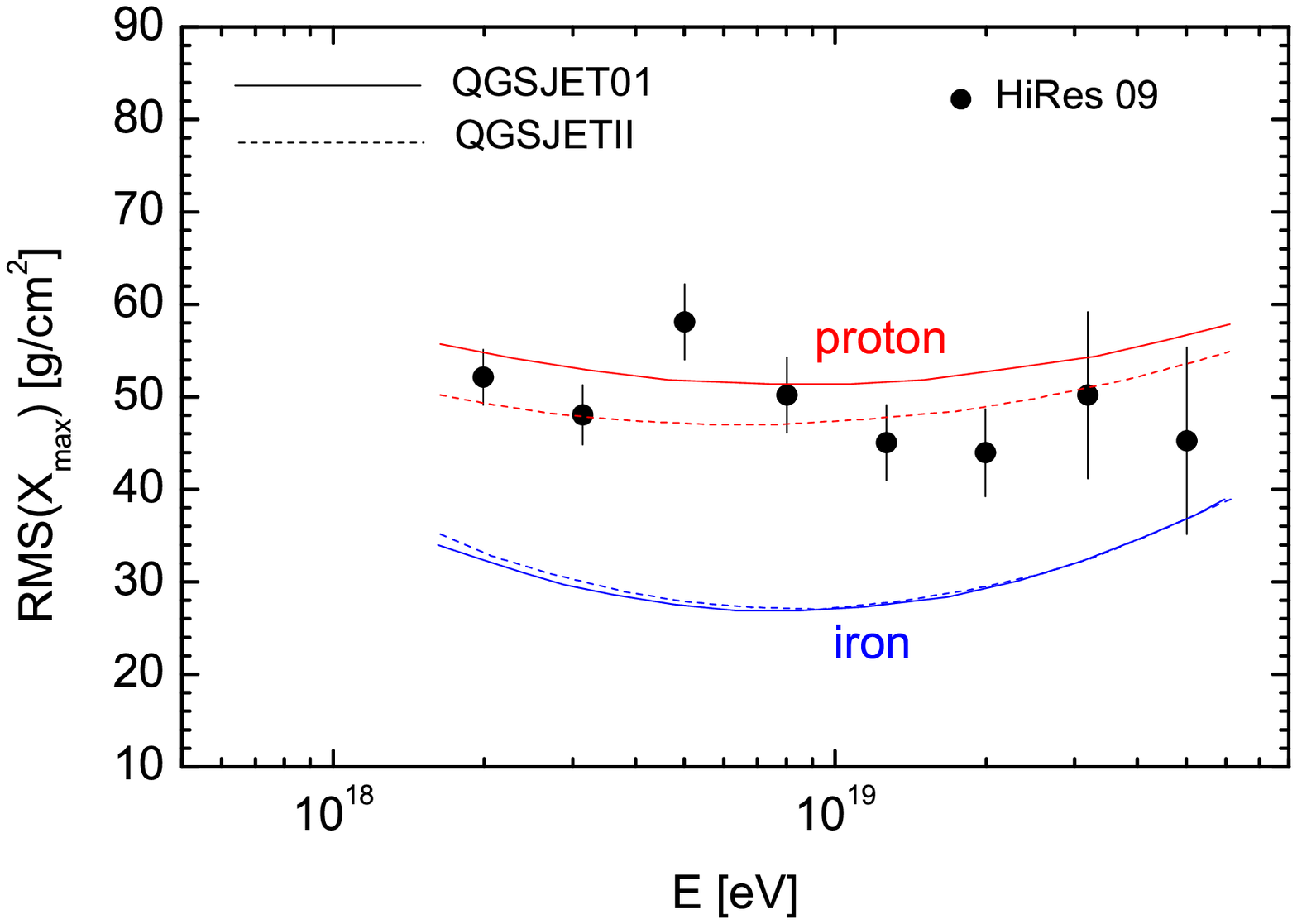}
 \end{minipage}
\medskip
 \begin{minipage}[ht]{62mm}
 \centering
 \includegraphics[width=62mm,height=50mm]{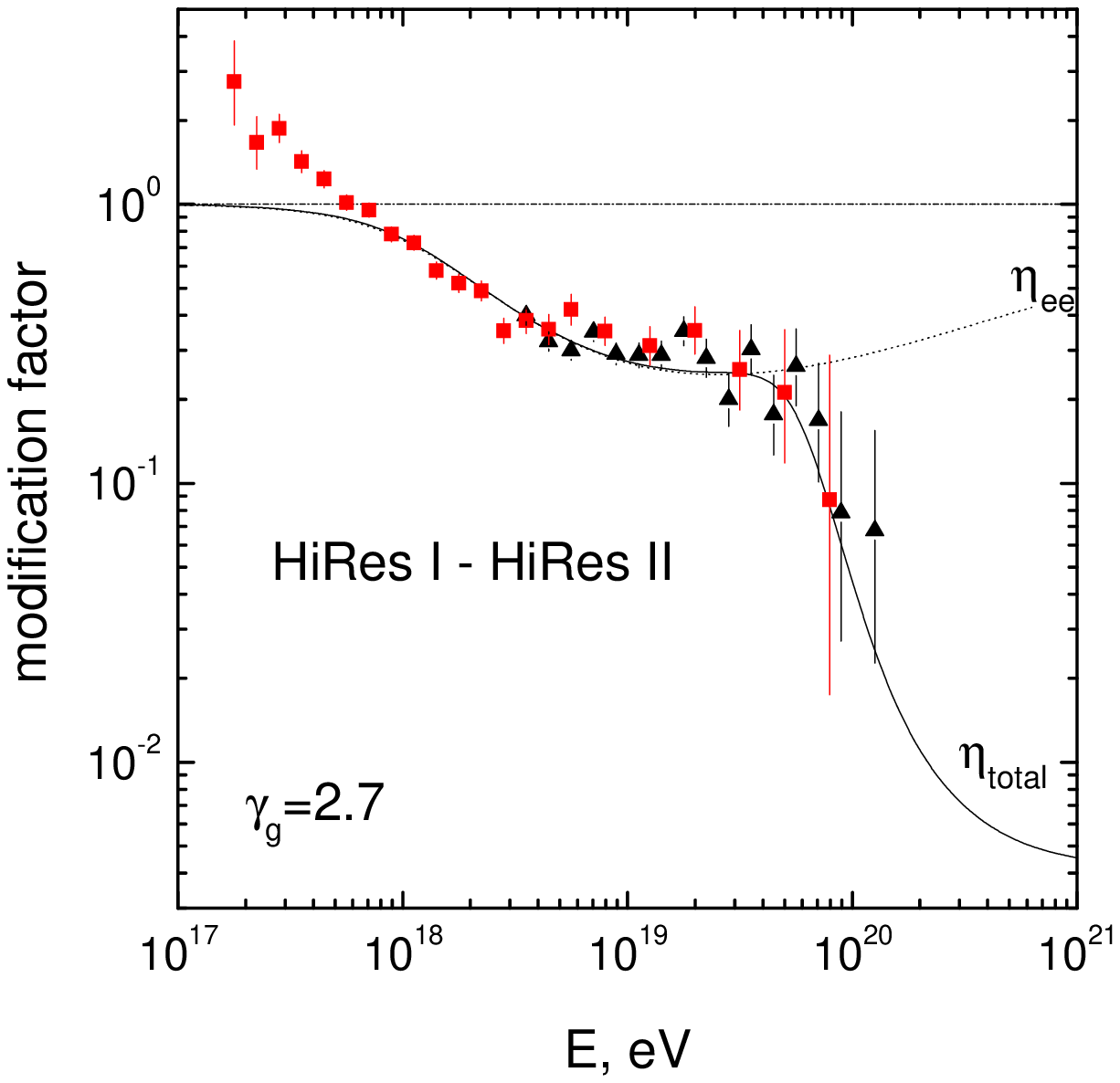}
 \end{minipage}
 \hspace{1mm}
 \vspace{-1mm}
 \begin{minipage}[h]{60 mm}
 \centering
 \includegraphics[width=58mm,height=50mm]{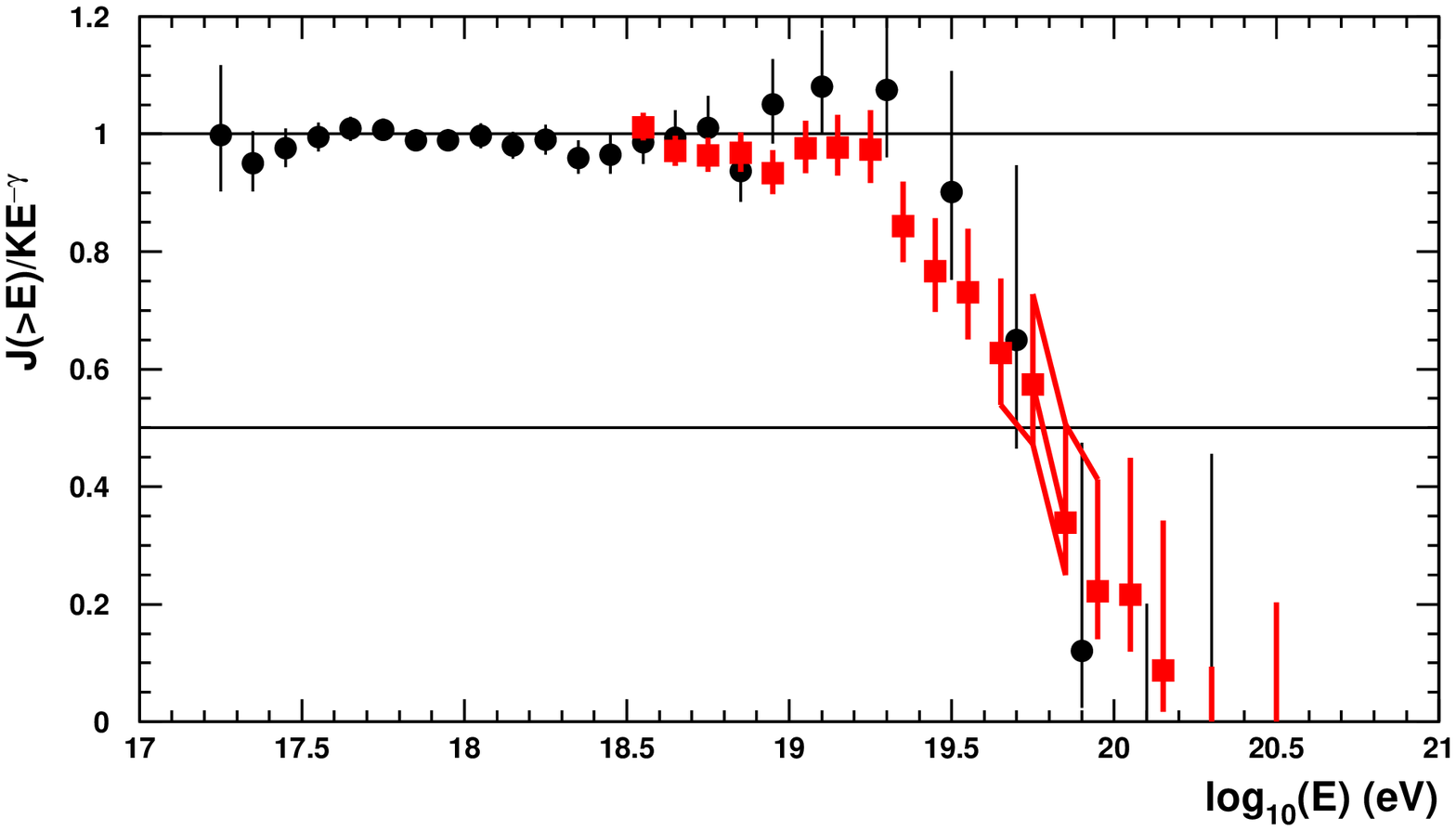}
 \end{minipage}
\end{center}
\vspace{-4 mm}%
\caption{ 
Mass composition and GZK cutoff as measured by the HiRes detector. In
two upper panels $\left\langle X_{\max}\right\rangle$ (left) and RMS (right) are presented as
function of the energy. Both agree with a pure proton composition,
shown by curves labeled 'proton'. The left-lower panel shows
differential energy spectrum in terms of the modification factor. One
can see a good agreement with the predicted shape of the GZK cutoff. The
right-lower panel shows the quantity $E_{1/2}$ in the integral
spectrum. This energy, a characteristic of the GZK cutoff, is found as
$E_{1/2}= 10^{19.73\pm 0.07}$~eV in good agreement with theoretical
prediction $E_{1/2}= 10^{19.72}$~eV (see the text).
} %
\label{fig:GZK-hires}
\end{figure*} %
\begin{equation} \eta_{\rm obs} \propto J_{\rm obs}(E)/E^{-\gamma_g}.
\label{obsMF}
\end{equation} 
Here $\gamma_g$ is the exponent of the generation spectrum $Q_{\rm
gen}(E_g) \propto E_g^{-\gamma_g}$ in terms of initial proton energies
$E_g$. Fig.~\ref{fig:dips} shows that both the pair production dip and
the beginning of the GZK cutoff up to $80$~EeV are well
confirmed by experimental data of Akeno-AGASA, HiRes, Yakutsk and TA
\cite{data}. 

The comparison of the theoretical dip with observational data includes 
only two free parameters: exponent of the power-law generation spectrum
$\gamma_g$ (the best fit corresponds to $\gamma_g=2.6 - 2.7$) and the
normalization constant to fit the $e^+e^-$-production dip to the
measured flux. The number of energy bins in the different experiments is
$20 - 22$. The fit is characterized by $\chi^2/{\rm d.o.f.} = 1.0 - 1.2$
for AGASA, HiRes and Yakutsk data. For this fit we used the modification
factor without cosmological evolution of sources. As was explained
above, using two additional evolutionary parameters, $m$ and 
$z_{\rm max}$, one can further improve the agreement.

In Fig.~\ref{fig:dips} one can see that at $E \lesssim 0.6$~EeV the
experimental modification factor, as measured by Akeno and HiRes,
exceeds the theoretical modification factor. Since by definition the
modification factor must be less than one, this excess signals the
appearance of a new component of cosmic rays at $E < E_{\rm tr} \approx
0.6$~EeV, which can be nothing else but the Galactic cosmic rays. This
interpretation is confirmed by the transition of a heavy component to
protons in the upper-left panel, that with good accuracy occurred at the
same energy. Thus, according to HiRes data the transition from
extragalactic to Galactic cosmic rays, occurs at energy $E_{\rm tr}
\sim 0.6$~EeV and is accomplished at $E \sim E_b \approx 1$~EeV (see 
left-upper panel in Fig.~\ref{fig:GZK-hires} as example).

\subsection{GZK cutoff.} 
\label{sec:GZK} 

The two largest Extensive Air Shower (EAS) detectors, HiRes and
Pierre Auger Observatory \cite{HiRes-GZK,Auger-GZK}, have observed a sharp
steepening in the UHECR spectrum at $E \gtrsim (30 - 50)$~EeV. Both
collaborations claimed that the observed steepening is consistent with the 
GZK cutoff. But as a matter of fact, there is a dramatic conflict between 
these two results, which still leaves the problem open.

In this subsection we analyze the data of HiRes which provide a strong 
evidence in favor of the GZK cutoff. These data are supported also by the 
TA observations \cite{TA-GZK}. The data of PAO will be considered in the 
next subsection.

To interpret convincingly the spectrum steepening as the GZK cutoff
one must prove that:
\begin{enumerate}
\item energy scale of the cutoff and its shape correspond to theoretical 
predictions, \item the measured chemical composition is strongly dominated 
by protons.
\end{enumerate} 
\begin{figure*}[ht]
\begin{center}
 \begin{minipage}[h]{70 mm}
 \centering
 \includegraphics[width=65 mm]{{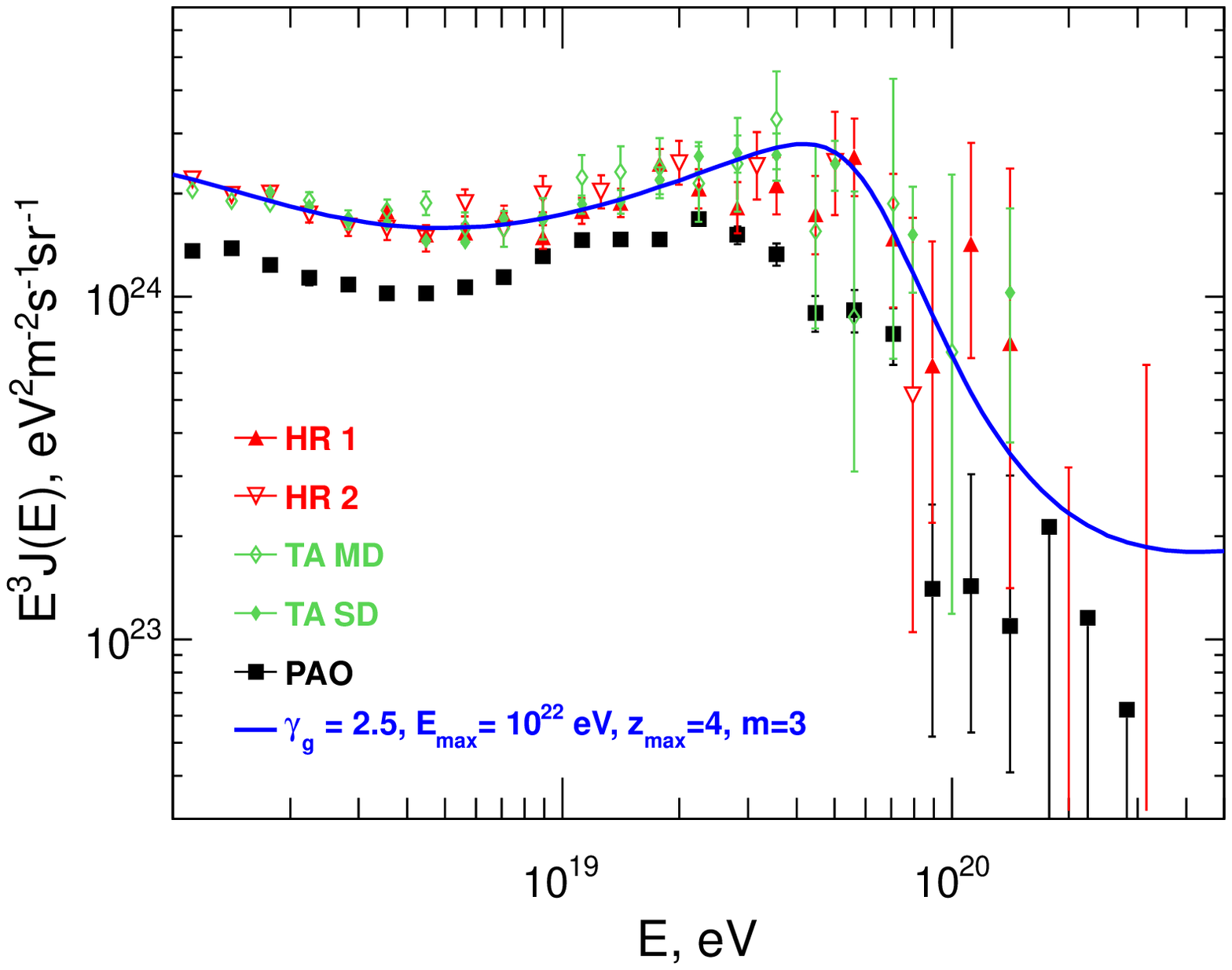}}
 \end{minipage}
 \hspace{1mm}
 \vspace{-1mm}
 \begin{minipage}[h]{70 mm}
 \centering
 \includegraphics[width=65 mm]{{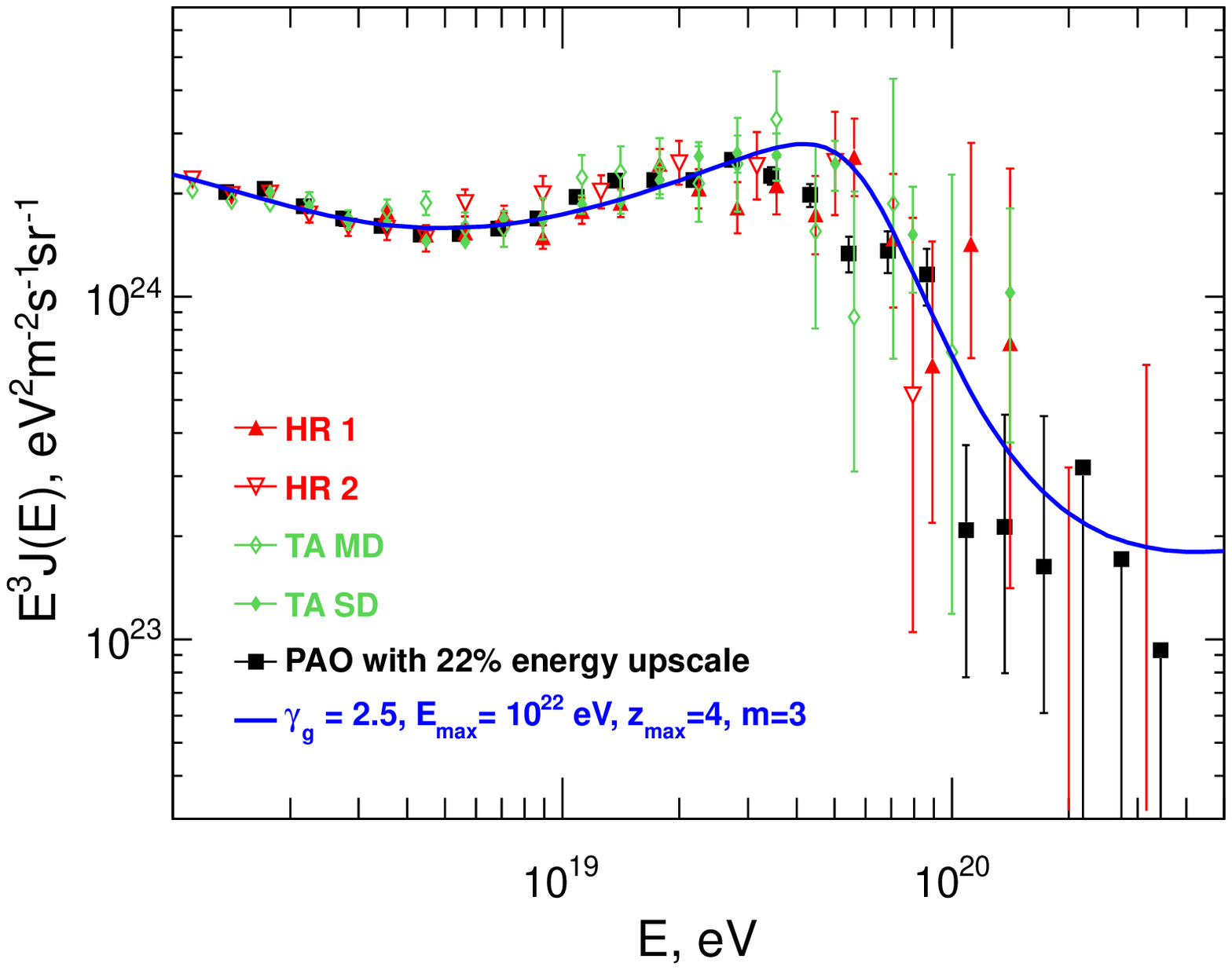}}
 \end{minipage}
 \end{center}
\vspace{-4 mm}%
\caption{
PAO spectrum \cite{PAO-11,Salamida} (filled squares) compared with 
 HiRes/TA spectra and with calculated spectrum (full line) with
 parameters indicated in the figures. In the {\em left panel} the 
 spectra are shown with energy scales as determined in the experiments. 
 In the {\em right  panel} the data of all three experiments are recalibrated using the  pair-production dip as a standard candle. For HiRes and TA the
 recalibration factor is 1.0, since these detectors describe well the
 dip. The PAO recalibration factor should be 1.22 to fit the dip shape.
 It is not trivial that after this recalibration the PAO flux coincides
 well with HiRes and TA fluxes. However, the contradiction of three
 PAO bins at $40 - 50$~EeV with the predicted GZK shape remains.
}
\label{fig:PAO-spct}
\end{figure*} %
\begin{figure*}[ht]
\begin{center}
 \begin{minipage}[ht]{60 mm}
 \centering
 \includegraphics[width=60 mm]{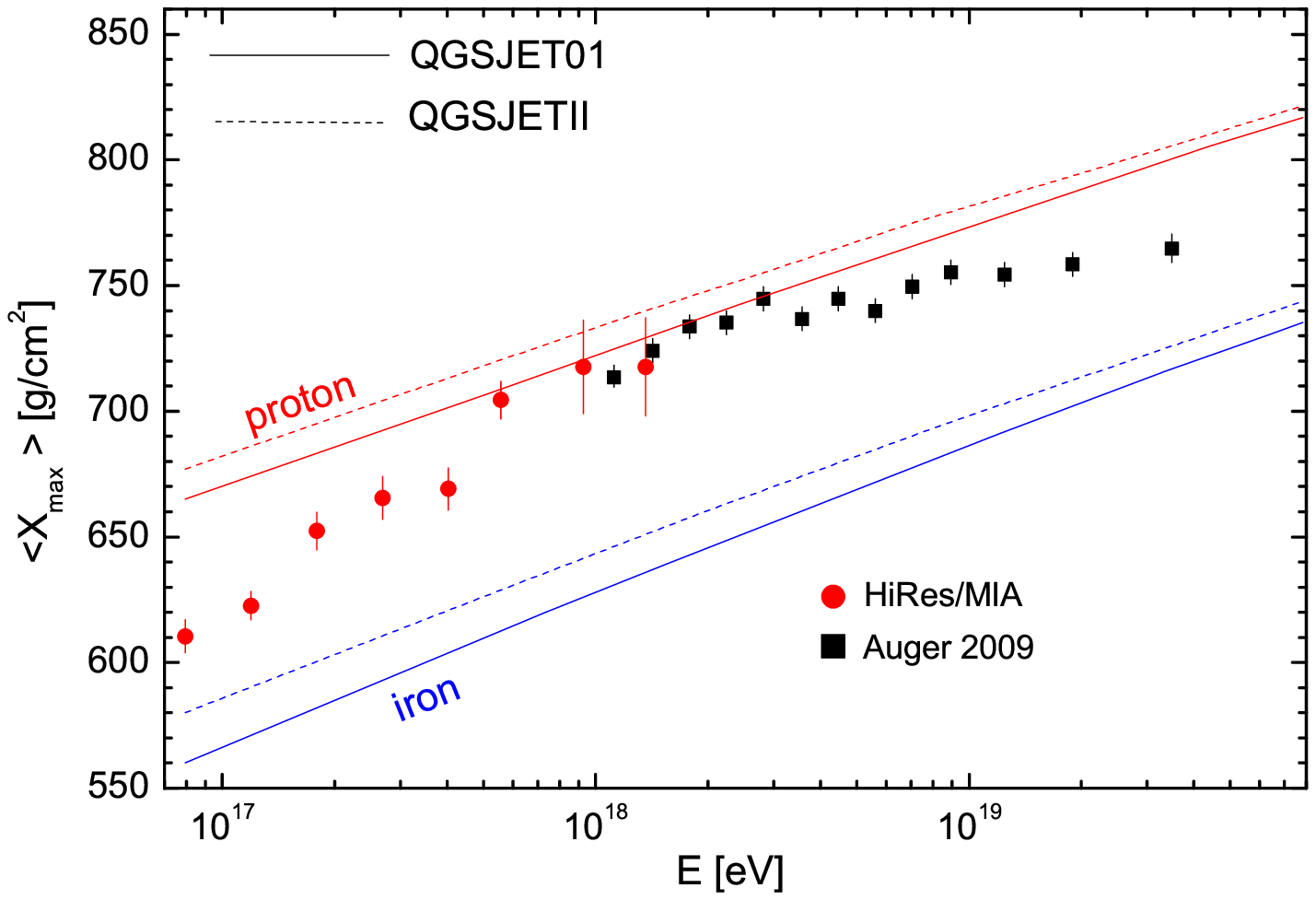}
 \end{minipage}
 \hspace{1mm}
 \vspace{-1mm}
 \begin{minipage}[h]{60 mm}
 \centering
 \includegraphics[width=60 mm]{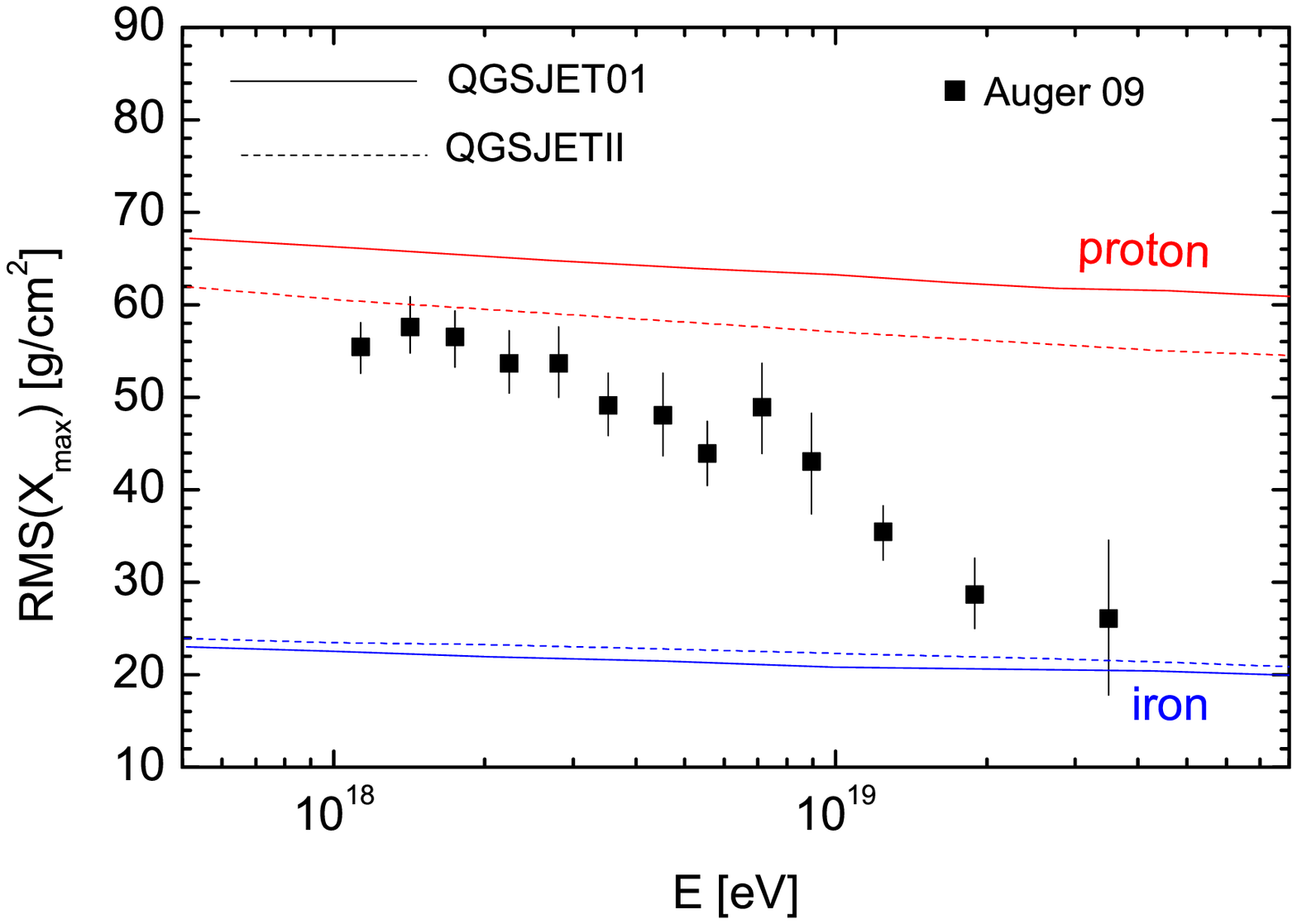}
 \end{minipage}
 \end{center}
\vspace{-4 mm}%
\caption{
Auger data \cite{Auger-mass} for $\left\langle X_{\max}\right\rangle$ as function of the energy (left panel) and for RMS($X_{\max}$), the width 
of the distribution over $X_{\max}$, (right panel). The calculated values 
for protons and Iron are given according to QGSJET01 \cite{QGSJET} and 
QGSJETII \cite{QGSJETII} models. One can see from the right panel that RMS
distribution becomes more narrow with increasing energy which implies a
progressively heavier mass composition. 
}
\label{fig:massAu}
\end{figure*} %
Mass composition of the CR spectrum in turn may be measured with the help of 
\begin{itemize}
\item $\left\langle X_{\rm max}\right\rangle(E)$, the
average depth of atmosphere in $g/cm^2$, where a shower with energy $E$
reaches its maximum, and
\item RMS($X_{\max}$), which is the width of the distribution over $X_{\max}$.
\end{itemize} 
These values measured by HiRes are displayed in 
Fig.~\ref{fig:GZK-hires}. From the left-upper panel of Fig.~\ref{fig:GZK-hires} 
one can see that the chemical composition changes from very heavy elements, 
probably Iron, at $E \sim 0.1$~EeV, (data of HiRes-MIA \cite{hires-mia}) to 
protons at $E\sim 1$~EeV (data of HiRes \cite{data}). Such evolution fully 
corresponds to the SM for GCR (see Fig.~\ref{fig:b-volk}). RMS$(X_{\max})$, 
a very sensitive tool for mass composition, also provides evidence for a 
proton-dominated composition at $E \gtrsim 1$~EeV and up to the highest 
energies (see upper-right panel of Fig.~\ref{fig:GZK-hires}). Differential 
energy spectrum of the GZK feature in the modification factor form
(left-lower panel) is in a reasonably good agreement with the theoretical 
prediction, though better statistics at higher energies is still needed for 
a final conclusion.

The {\em integral energy spectrum} of UHE protons, $J_p(>E)$, has
another specific characteristic of the GZK cutoff, the energy $E_{1/2}$
\cite{BG88}. It is based on the observation that the calculated 
integral spectrum below $50$~EeV is well approximated by a power-law
function: $J_p(>E) \propto E^{-\tilde{\gamma}}$. At high energy this
spectrum steepens due to the GZK effect. The energy where this steep
part of the spectrum equals to half of its power-law extrapolation,
$J_p(>E)=K E^{-\tilde{\gamma}}$, defines the value of $E_{1/2}$. This
quantity is found to be practically model-independent; it equals to
$E_{1/2} = 10^{19.72}\mbox{ eV} \approx 52.5 $ EeV \cite{BG88}.
Fig.~\ref{fig:GZK-hires} demonstrates how the HiRes collaboration found
$E_{1/2}$ from observational data \cite{E_1/2hires}. The ratio of the
measured integral spectrum $J(>E)$ and the low-energy power-law
approximation $KE^{-\tilde{\gamma}}$ was plotted as a function of
energy. This ratio is practically constant in the energy interval $0.3 -
40$~EeV, indicating that the power-law approximation is a good fit,
indeed. At higher energy the ratio falls down and intersects the
horizontal line $0.5$ at the energy defined as $E_{1/2}$. It results in
$E_{1/2}= 10^{19.73\pm 0.07}$~eV, in excellent agreement with the
predicted value.

Thus, one may conclude that the HiRes data presented in
Fig.~\ref{fig:GZK-hires} indicate the proton-dominated chemical
composition and the presence of the GZK cutoff in both differential and
integral spectra. The conclusion about proton composition is further
supported by the recent TA data \cite{data}. 

\subsection{PAO data: energy spectrum and mass composition} 
\label{sec:auger} 

The spectrum of the Pierre Auger Observatory, shown by filled squares in
the left panel of Fig.~\ref{fig:PAO-spct}, differs from the HiRes and 
Telescope Array spectra both in absolute normalization and in shape. 
However, one is allowed to shift the energy scale within systematic
energy errors. Since the energy position of the pair-production dip is 
rigidly fixed by the interaction with CMB, in Ref.~\cite{BGGprd} and 
\cite{Aletal} it was proposed to use the dip as an energy calibrator 
\begin{figure*}[ht]
\begin{center}
 \begin{minipage}[ht]{60 mm}
 \centering
 \includegraphics[width=60 mm]{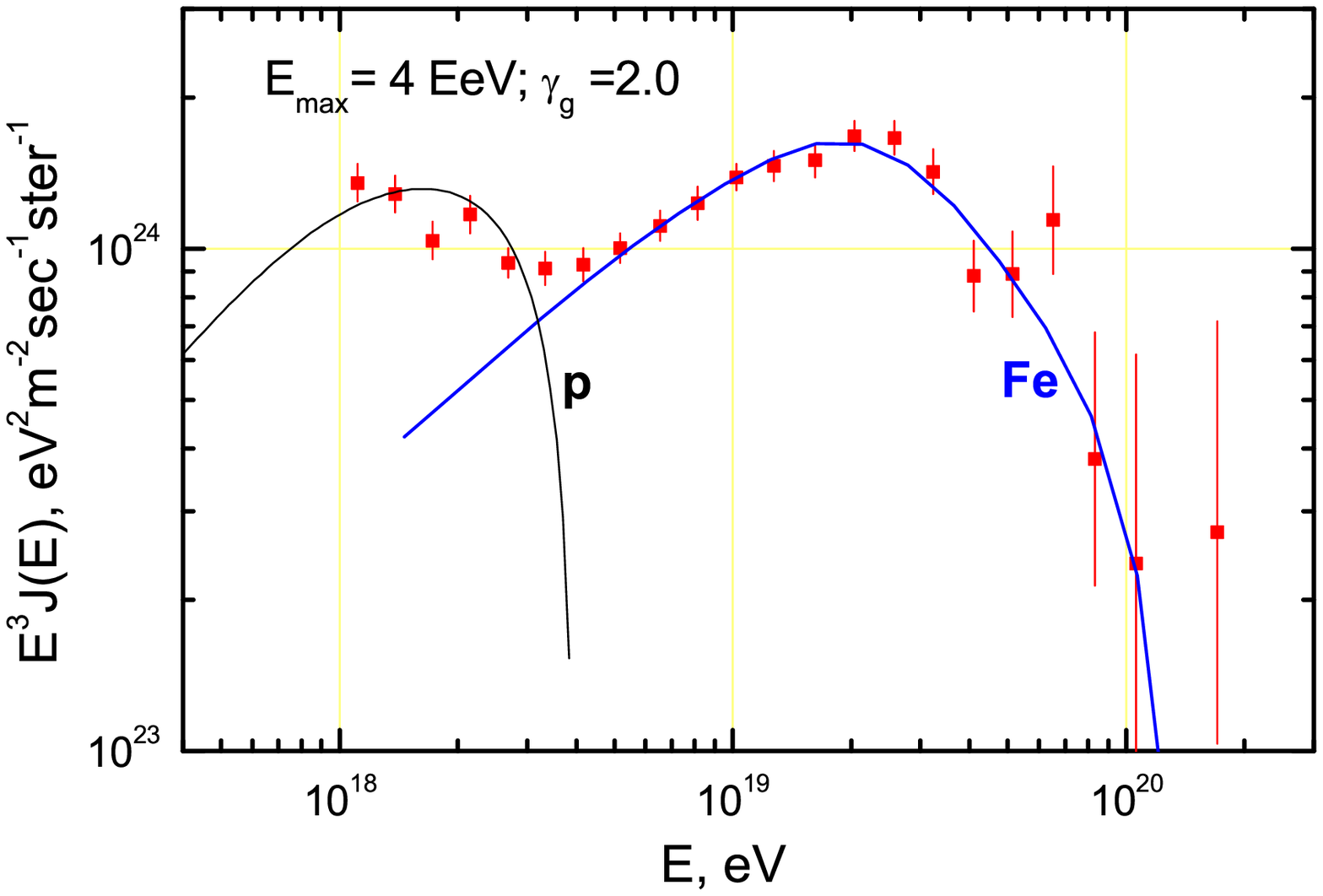}
 \end{minipage}
 \hspace{1mm}
 \vspace{-1mm}
 \begin{minipage}[h]{60 mm}
 \centering
 \includegraphics[width=60 mm]{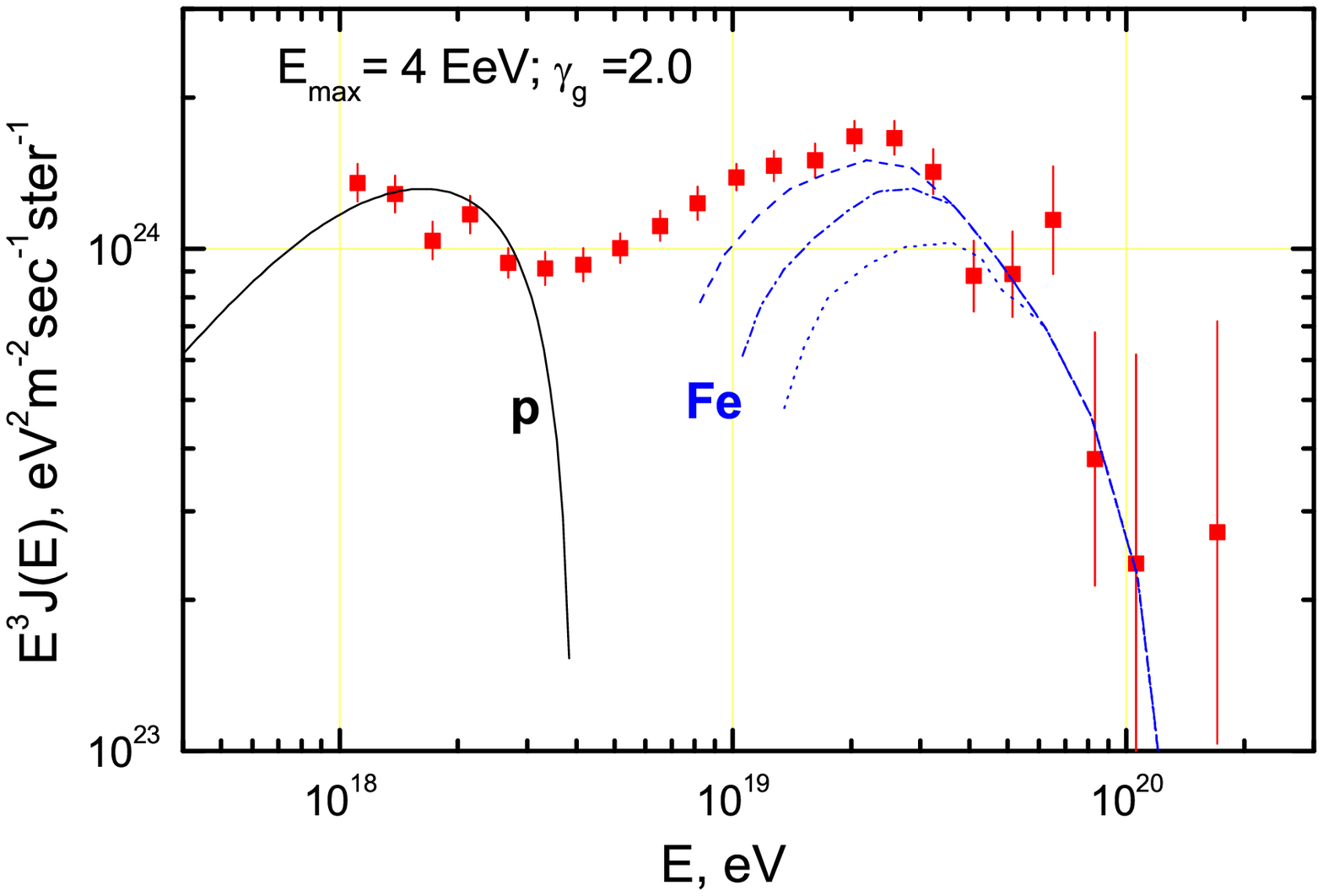}
 \end{minipage}
 \end{center}
\vspace{-4 mm}%
\caption{
{\em Left panel:} Energy spectrum in the two-component model with
protons and Iron nuclei for homogeneous distribution of sources and
with $\gamma_g=2.0$ and $E_{\max}=4$~EeV. {\em Right panel:} As in
left panel but with a diffusion cutoff. The gap is expected to be filled
by intermediate mass nuclei. 
}
\label{fig:disapp-spectrum}
\end{figure*} %
('standard candle'). The data of HiRes and
TA have recalibration factors $k_{\rm cal}=1.0$, since these data agree
well with the energy position and shape of the pair-production dip (see
left panel of Fig.~\ref{fig:PAO-spct}). The Auger data (left panel of
Fig.~\ref{fig:PAO-spct}) disagree with the dip, but with the recalibration
factor $k_{\rm cal} \simeq 1.22$ a reasonable agreement can be reached. It 
is far non-trivial that simultaneously the Auger flux agrees well with 
the fluxes of HiRes and TA. This procedure was applied in
\cite{BGGprd,Aletal} imposing the condition of minimizing $\chi^2$ after
recalibration of the data of AGASA, HiRes and Yakutsk detectors. A 
similar procedure based on the dip energy calibration was also used by 
Bl\"umer et al.\ \cite{GZK-rev} for all experiments with recalibration 
factors $1.0$, $1.2$, $0.75$, and $0.625$ for the HiRes, PAO, AGASA and
Yakutsk detectors, respectively; a good agreement for all fluxes was 
demonstrated. This procedure was also successfully applied in
\cite{Yak}. 

The apparent coincidence of the PAO and HiRes/TA spectra is related to the 
low energy part of the energy spectrum in Fig.~\ref{fig:PAO-spct}. At
higher energies statistical uncertainties are too large to distinguish
between the spectra. 

While the HiRes and TA spectra are compatible with the GZK cutoff, the Auger 
spectrum is not. The steepening in the upscaled PAO
spectrum  starts at energy $E \lesssim 40$ EeV, lower than $E_{\rm GZK}
\simeq 50$ EeV, and in three successive energy bins the PAO flux is 
significantly lower than one predicted for the GZK shape as shown in 
the right panel of Fig.~\ref{fig:PAO-spct}. We could not reconcile 
the PAO cutoff shape with the GZK behavior by including in 
calculations different generation indices $\gamma_g$, evolution regimes,
low acceleration maximum energy $E_{\max}$,  local over-density of
sources etc.  

This disagreement is quite natural taking into account the mass composition 
measured by PAO (see Fig.~\ref{fig:massAu}), which dramatically differs from 
that of HiRes shown in Fig.~\ref{fig:GZK-hires}. The dependence of 
the RMS($X_{\max}$) on energy, which steadily 
decreases and approaches the Iron value at $E \approx 35$~EeV, is a 
strong example of this disagreement. Low RMS, i.e.\ small fluctuations, 
is a typical and reliable feature of a heavy nuclei composition. 
These data are further strengthened by other PAO measurements provided
by surface detectors. They allow to extract two mass-composition 
dependent quantities: the atmospheric depth $\left\langle X_{\max}^{\mu}\right\rangle$, where 
muon-production rate reaches maximum, and maximum zenith angle 
$\theta_{\max}$ determined by the signal rise-time in surface 
Cerenkov detectors. Measurements of both quantities confirm the 
heavy mass composition and its dependence on energy obtained
with the help of $\left\langle X_{\max}\right\rangle(E)$ and RMS($X_{\max}$) 
\cite{Garcia-GamezICRC} and \cite{Gazon2012}. The soon expected data 
on muon flux from the Auger Muon and Infill Ground Array (AMIGA) \cite{AMIGA} 
will further clarify the mass composition. 

An early steepening of the spectrum observed by the PAO can be explained 
as a property of nuclei propagation. However, it is a problem to 
explain simultaneously both spectrum and mass composition 
of PAO. 

A wide class of {\em mixed composition} models \cite{mixed} fails to fit
the PAO mass composition at the highest energies. Most of them assume a
source mass composition close to Galactic CRs, i.e.\ enhanced by
protons. The observed mass composition in these models becomes heavier
only at $E \gtrsim 50$~EeV, when protons disappear due to the GZK cutoff
\cite{mixed}. In the paper \cite{mixed}, by Allard et al. (2008), an 
exceptional case of primaries being pure Iron nuclei was considered, 
though authors refer to this possibility as to an unnatural one. This model 
(see also \cite{AllardReview} fig. 4 and \cite{Taylor11}) explains the observed 
Auger energy spectrum.

Mass composition becoming progressively heavier with increasing energy 
appears in the rigidity-dependent acceleration, when the maximum
attainable energy for a nucleus with charge number $Z_i$ is
$E_i^{\max}=Z_i E_p^{\max}$ and the contribution of nuclei with charge
$Z'<Z$ to this energy disappears, while heavier nuclei with larger $Z$
exist. In this way the 'disappointing model' \cite{disapp} was built. 

The disappointing model is based on the following observations and
assumptions: 
\begin{itemize}
\item According to the HiRes (Fig.~\ref{fig:GZK-hires}) and PAO 
(Fig.~\ref{fig:massAu}) data, the observed primaries at energy 
$(1 - 3)$~EeV are predominantly protons. 
\item It is assumed that they have extragalactic origin.
\item The particles at higher energies are extragalactic nuclei with charge 
number $Z$ steadily increasing with energy.
\item Acceleration of CRs in sources is rigidity dependent, 
$E_i^{\max}=Z_i \times E_p^{\max}$.
\end{itemize}

It was demonstrated in \cite{disapp} that to avoid a proton dominance at
the highest energies, one must assume that the maximum energy of the
accelerated protons is limited, $E_p^{\max} \lesssim (4 - 10)$~EeV.
This conclusion is valid for a large range of generation indices,
$\gamma_g \sim 2.0 - 2.8$, and for a wide range of cosmological
evolution parameters. The calculated proton and nuclei energy spectra
for $\gamma_g=2.0$, $E_{\max}=Z\times 4$~EeV and without cosmological
evolution are shown in Fig.~\ref{fig:disapp-spectrum}. In the left panel
the two component model (protons and Iron) is presented. In the right
panel of Fig.~\ref{fig:disapp-spectrum}
\begin{figure*}[t]
\begin{center}
\includegraphics [width=0.8\textwidth]{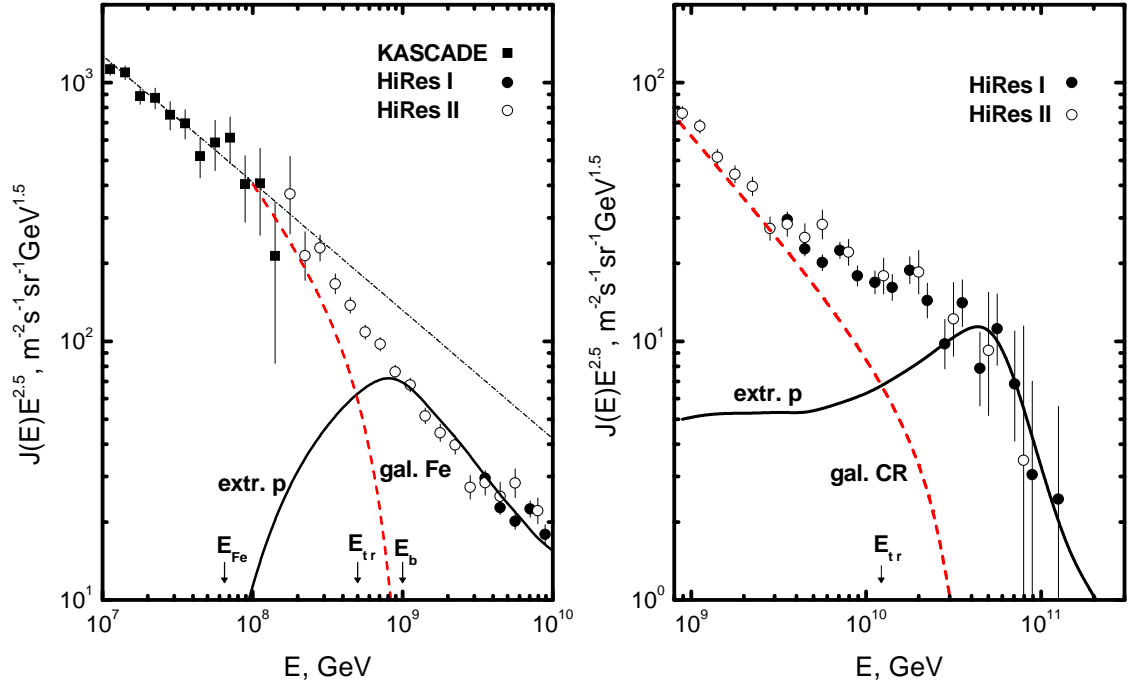}
\end{center}
\caption{
Transition in the dip (left panel) and ankle (right panel) models. In
 both cases solid lines give the calculated extragalactic proton
 spectrum and dashed lines describe the Galactic iron spectrum. $E_{\rm
 tr}$ is the energy of intersection of the Galactic and extragalactic
 spectra and $E_{\rm Fe}$ gives the position of the iron knee. $E_b
 \approx 1$~EeV in the left panel is the energy where the transition
 from Galactic to extragalactic CRs is completed.
}
\label{fig:dip-ankle}
\end{figure*}
intermediate primary nuclei are included in the framework of the
diffusive propagation through intergalactic magnetic fields (see
\cite{disapp} for details). The rigidity dependent acceleration,
$E_i^{\max}=Z_i\times E_p^{\max}$, provides a steadily increasing charge
number $Z$ as energy increases. 

The disappointing model as it is presented here and in \cite{disapp}
is not complete. The energy spectra must be calculated in diffusive
approximation for different distances between sources. Energy
dependence of $\left\langle X_{\max}\right\rangle(E)$ and RMS($X_{\max}$) 
must be consistent with calculated energy spectrum.

This model was called \emph{'disappointing'} because of lack of many 
signatures predicted in proton-dominated models, such as cosmogenic 
neutrino production and correlation of CR arrival directions with 
distant sources.

Recently a similar model \cite{parizot}, based on the rigidity-dependent
acceleration, was proposed for making the chemical composition heavier
with energy increasing. This model uses the reasonable assumption
\cite{KS,Aletal} that the space density $n_s$ of UHECR sources decreases
with increasing of the maximum acceleration energy. The idea may be
explained in the following way. 

Consider accelerators which produce protons ($p$) and nuclei ($i$) with
maximum energy $E_p^{\max}$ and $E_i^{\max} = Z_i\times E_p^{\max}$, 
respectively. The space density of sources is assumed to depend on
$E_{\max}$, 
\[
n_s(E_p^{\max}) \propto (E_p^{\max})^{-\beta}.
\] 
Generation rate of protons with energy $E$ per unit space volume
is proportional to
\begin{equation}
Q_p(E) \propto \int_E dE_p^{\max} (E_p^{\max})^{-\beta} 
\propto E^{-\beta+1}, 
\label{eq:Q_p}
\end{equation}
because sources with $E_p^{\max} < E$ do not contribute. For nuclei
$i$
\begin{equation}
Q_i(E) \propto \int_{E/Z_i} dE_p^{\max} (E_p^{\max})^{-\beta} 
\propto Z_i^{\beta-1} E^{-\beta+1 }
\label{eq:Q_i}
\end{equation}
(sources with $E_p^{\max} < E/Z_i$, i.e.\ $Z_i E_p^{\max} < E$,
do not contribute). Then from Eqs.~(\ref{eq:Q_p}) and (\ref{eq:Q_i})
\begin{equation}
Q_i(E)/Q_p(E) \propto Z_i^{\beta - 1}.
\label{Q_i/Q_p}
\end{equation}
Eq.~(\ref{Q_i/Q_p}) shows that the production of nuclei is enhanced by
a factor $Z_i^{\beta -1}$; it can affect the mass composition
calculations in the mixed composition models. 

\section{Three transition models}
\label{transition}
In this section we discuss three models of transition from Galactic to
extragalactic CRs: {\em ankle}, {\em dip} and {\em mixed composition}
models. One feature is common for all of them: the transition is
described as an intersection of a steep Galactic spectrum with a more
flat extragalactic one. The agreement with the SM for GCR (see subsection
\ref{SM}) is one more criterion which these models have to respect. 
According to the Standard Model, the benchmark of the end of GCR is
given by the Iron knee at energy $E_{\rm Fe} \approx 80$~PeV (see
Fig.~\ref{fig:b-volk}) and at $E > E_{\rm Fe}$ it has an exponential cutoff. 

Motivated by the interpretation of the ankle as the transition to 
extragalactic CR at $E_a^{\rm tr} \sim (3 - 10)$~EeV, one has to assume
\cite{hillas2005,hillas2006,gaisser} an additional component of GCR 
\begin{figure*}[t]
  \vspace{5mm}
  \begin{center}
   \begin{minipage}[ht]{75 mm}
     \centering
     \includegraphics[width=\textwidth]{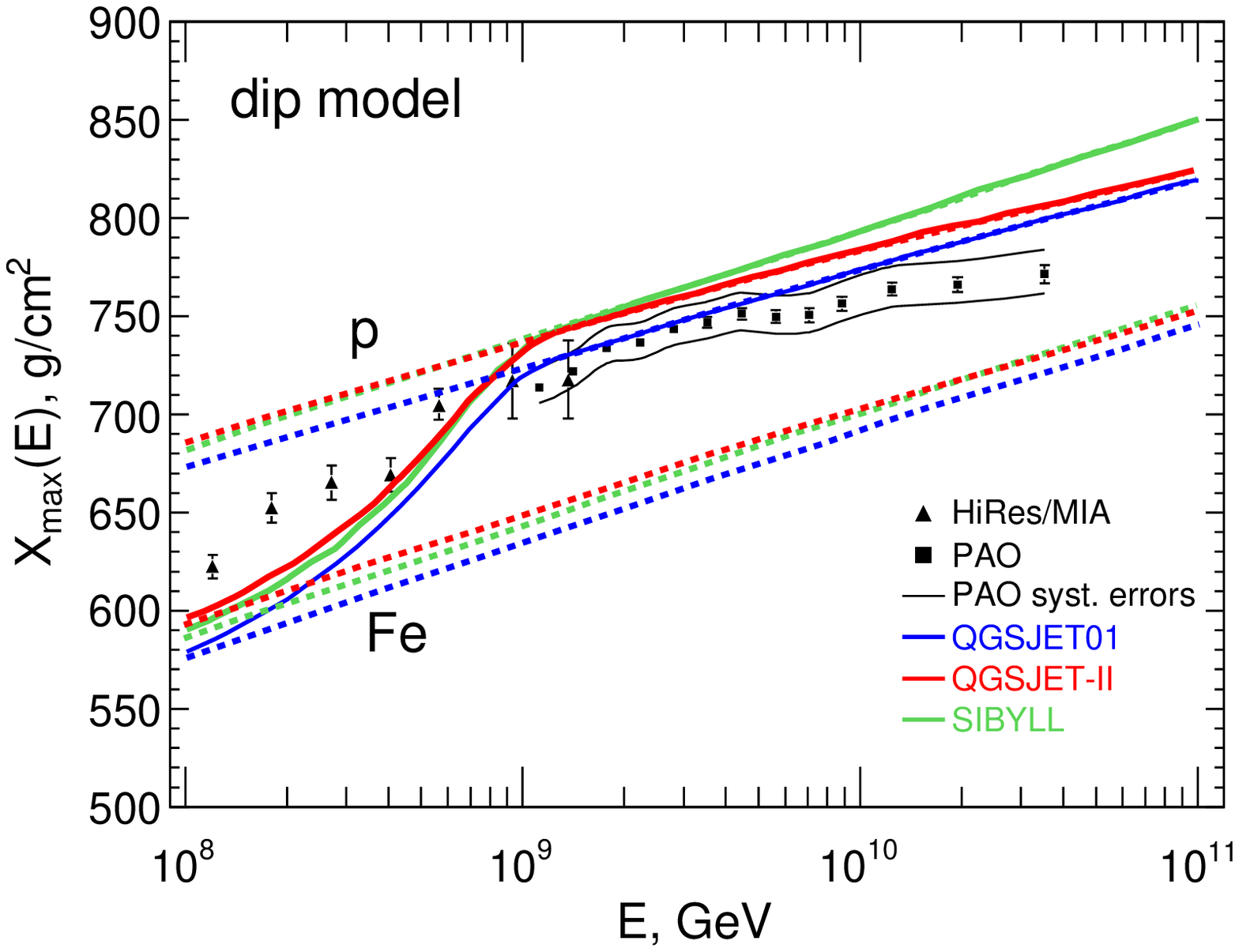}
   \end{minipage}
   \hspace{5mm}
   \begin{minipage}[ht]{75 mm}
     \centering
     \includegraphics[width=\textwidth]{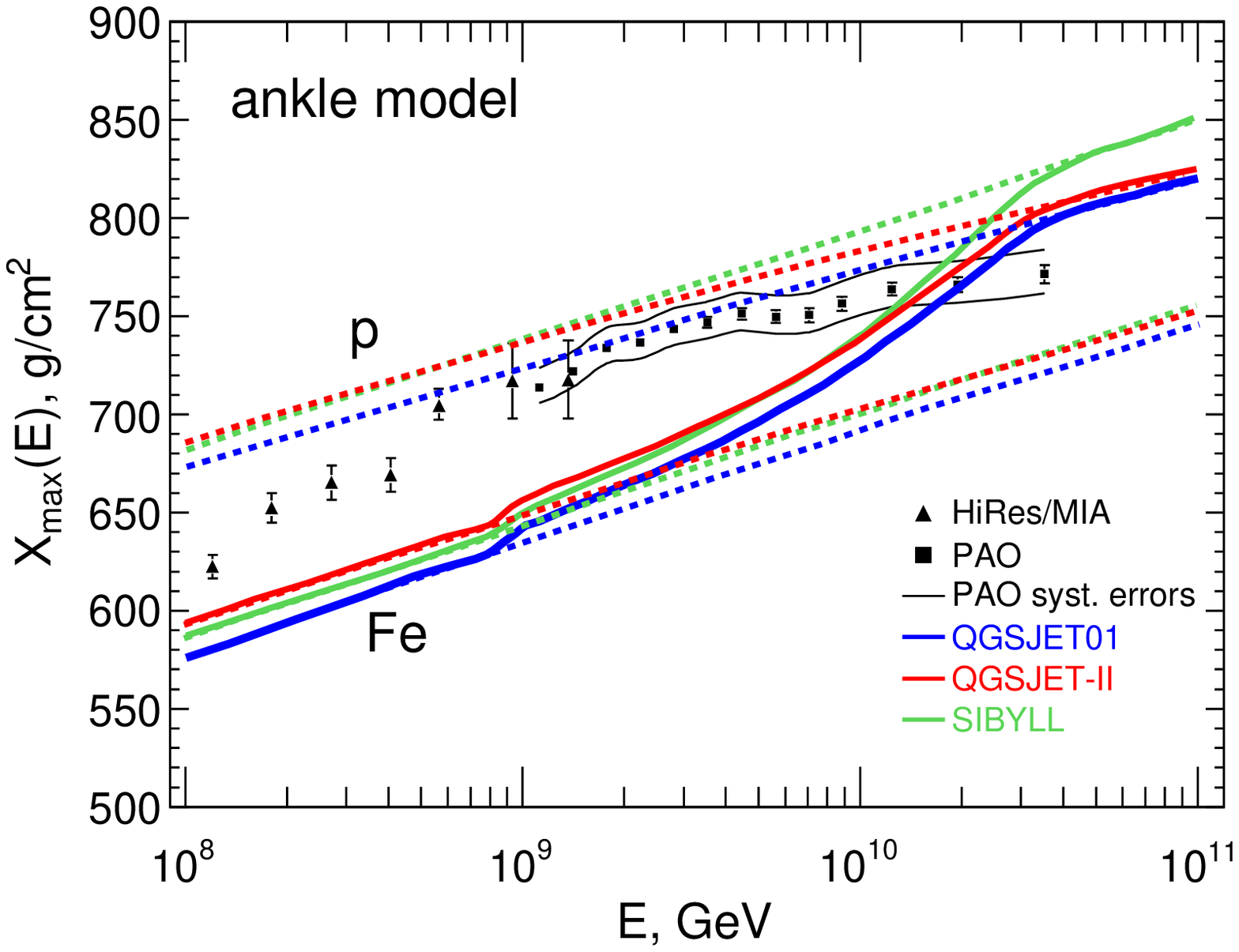}
   \end{minipage}
 
   \begin{minipage}[ht]{75 mm}
     \centering
     \includegraphics[width=\textwidth]{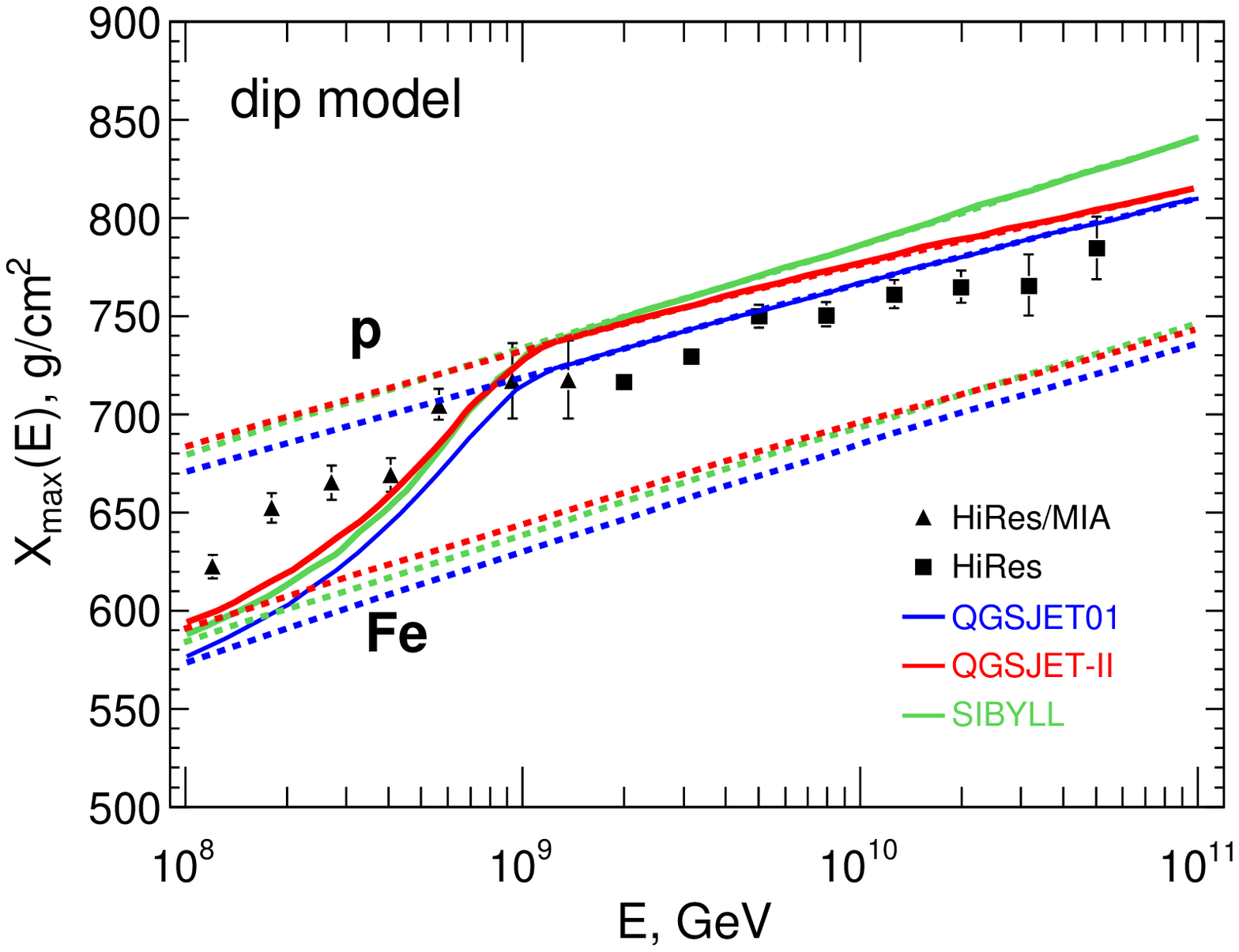}
   \end{minipage}
   \hspace{5mm}
   \begin{minipage}[ht]{75 mm}
     \centering
     \includegraphics[width=\textwidth]{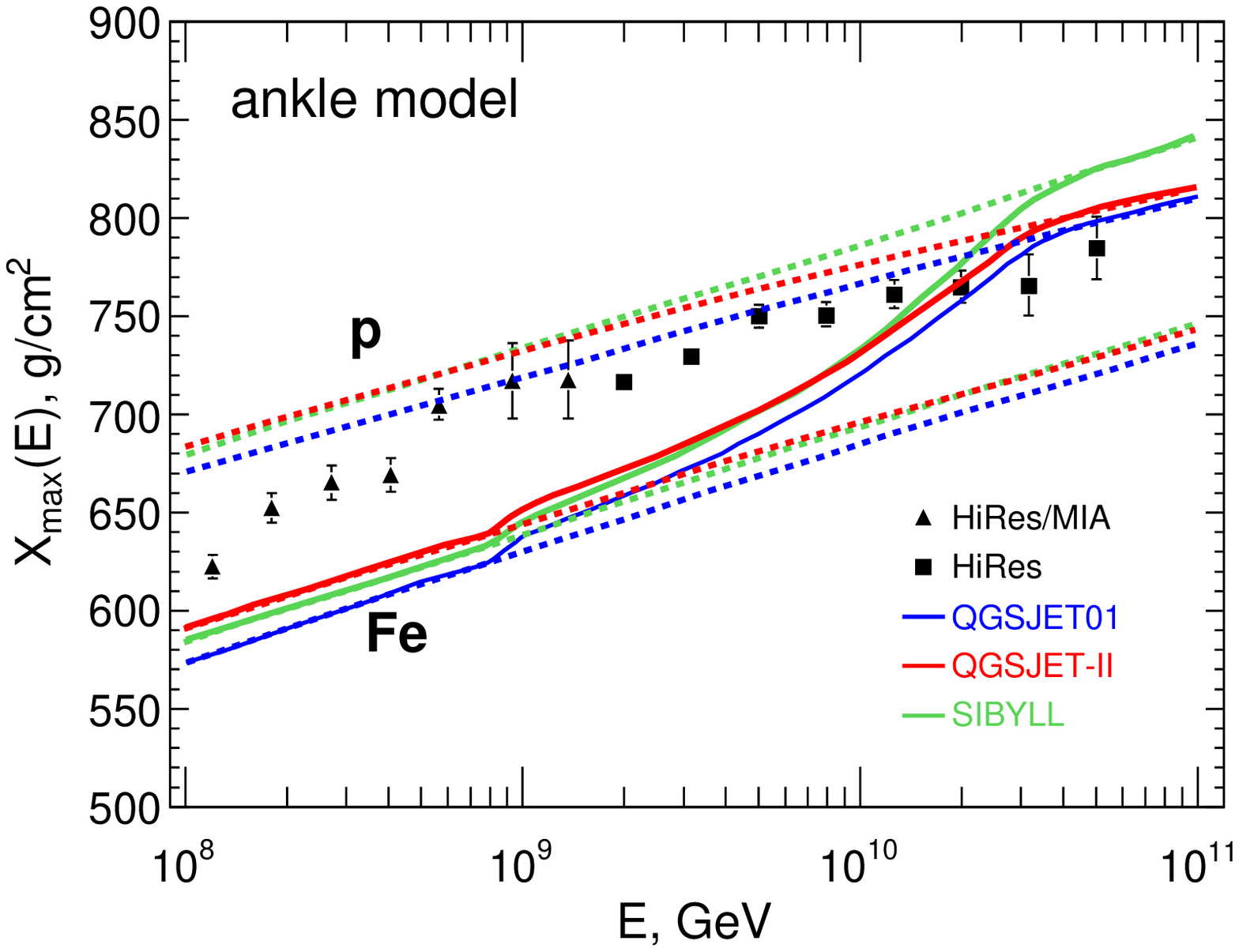}
   \end{minipage}

  \end{center}
\caption[ ]{
The calculated elongation curves $\left\langle X_{\max}\right\rangle(E)$ 
for the dip model (two left panels) and ankle model (right panels). The  calculated 
curves $\left\langle X_{\max}\right\rangle(E)$ are shown by 
the thick solid lines for QGSJET01 \cite{QGSJET}  model of interaction, 
by the thick dashed lines for QGSJET-II \cite{QGSJETII}, and by dotted lines for  
SIBYLL \cite{SIBYLL}. The data points are measurements of HiRes-Mia 
(filled triangles), HiRes and PAO (both filled boxes). The PAO data 
in upper panels with systematic errors, shown by the thin curves, are 
taken from \cite{PAO-11cc}. The lines 'p' and 'Fe' present the
elongation curves for proton and Iron which are used for calculations
of the model elongation curves in each panel. For PAO data they are 
theoretical curves, for HiRes they include cuts, detector's properties 
etc. and are taken from curves 'iron' and 'proton' in the upper-left panel 
of Fig.~\ref{fig:GZK-hires}. As main result of these 
plots one may notice the great discrepancy of the ankle model with the data.}

\label{fig:Xmax}
\end{figure*} %
accelerated up to energies much beyond the Iron knee (see subsection
\ref{NewComponent}). 

Observational data which have the power either to confirm or to reject
each transition model include energy spectrum, elongation curve
$\left\langle X_{\max}\right\rangle(E)$, RMS($X_{\max}$) and anisotropy. 
Below these models will be discussed in the historical succession of appearance: 
ankle, dip and mixed composition. 

\subsection{Ankle model}
\label{AnkleModel}
This is the traditional model based on the interpretation of the ankle
as the spectrum feature where transition occurs (see e.g.\ 
\cite{stanev2005} - \cite{waxman}). In fact, this is a very natural 
model since transition occurs because the extragalactic component is
very hard. This component is assumed to have a pure proton composition
with a flat generation spectrum $Q_{\rm extr.p} \propto E^{-2}$ valid
for non-relativistic shock acceleration. Energy losses modify the
spectrum insignificantly at $E \lesssim 40$~EeV. The beginning of the
ankle at $E_a^{\rm obs} \sim (3 - 4)$~EeV corresponds to the energy 
where fluxes of Galactic and extragalactic CRs get equal. Thus, the 
Galactic CRs should be presented by an additional component accelerated 
up to energy at least by factor $30 - 40$ times higher than the maximum 
energy in the Standard Model. In the majority of the ankle models, e.g.\  
\cite{hillas2005,hillas2006,WW2005,waxman}, the large fraction of the
observed cosmic rays has a Galactic origin at $E \gtrsim 10$~EeV.
To facilitate the acceleration problem one should assume a heavy-nuclei
composition of the new component. 

The transition at the ankle is illustrated by the right panel of 
Fig.~\ref{fig:dip-ankle}. The curve ``extr.p'' presents the calculated 
extragalactic flux of protons and the dash-dot line gives the Galactic
CR spectrum. The latter is obtained by subtracting the extragalactic
component from the total observed flux following the procedure first
suggested in \cite{BGH}. The observed dip in the spectrum may be
explained by the Hill-Schramm's mechanism \cite{HS85}. 

Another problem of the ankle model is the contradiction with the
measured average depth of EAS maximum, $\left\langle X_{\max}\right\rangle(E)$, in the energy
range $(1 - 5)$~EeV. While all data, including both HiRes and PAO, 
show proton or light nuclei composition here, the ankle model 
needs a heavy Galactic component, predicting too small 
$\left\langle X_{\max}\right\rangle(E)$ in contradiction with observations (see the right 
panel of Fig.~\ref{fig:Xmax} and right panel of Fig.~4 in 
\cite{ABBO}). This contradiction is found also in \cite{mixed}.

The ankle at an energy 
higher than $3$~EeV contradicts also the anisotropy calculated in
\cite{anis-ankle}.
\vspace{10mm}

\subsection{Dip model}
\label{DipModel} 

The dip model is based on the assumption that UHECR at $E \gtrsim 1$
EeV are mostly extragalactic protons. This assumption is confirmed by
the observation of the {\em pair-production dip} in the energy range
$(1 - 40)$~EeV and the beginning of the GZK cutoff in the energy range
$(40 - 100)$~EeV (see Fig.~\ref{fig:dips}). Both features are
signatures of a proton dominated spectrum. As was discussed above the 
shape of the dip allows an admixture of light nuclei, though not more 
than $15\%$. The transition from Galactic to extragalactic CRs occurs 
as the intersection at $E_{\rm tr} \sim 0.5$~EeV of the steep Galactic
component (dashed line in the left panel of Fig.~\ref{fig:dip-ankle})
with the flat extragalactic proton component shown by 'extr.p' curve
(this curve appears to be falling down since the spectrum is multiplied by
$E^{2.5}$). The flatness of the extragalactic spectrum is provided by
the distribution of sources over the maximum acceleration energy $n_s
(E_{\max}) \propto E_{\max}^{-\beta}$ (see below) or, in the case of
diffusive propagation, by the 'magnetic flattening'
\cite{Aletal,Lem2005,AB2005}. The transition is completed at energy $E_b
\approx 1$~EeV, i.e.\ it occurs at a visible feature in the CR spectrum
known as the 'second knee', at energy between $E_{\rm tr}$ and $E_b$ in 
the left panel of Fig.~\ref{fig:dip-ankle}.

The basic features of the {\em dip model} are as follows
\cite{BGGPL,BGGprd,Aletal}:
\begin{itemize}
\item The primary flux is strongly proton-dominated.
\item Sources are e.g.\ AGN \cite{BGG-AGN} with a neutron 
mechanism for particle escape \cite{berez77,dermerAGN} which provides a
pure proton generation spectrum. Another case of enhancement in the proton 
fraction at relativistic shock acceleration is given in \cite{Aletal}. 
\item To reproduce the observed shape of the dip, indices for the 
generation rate per unit co-moving volume $Q(E) \propto E^{-\gamma_g}$
in models without evolution are to be $\gamma_g=2.6 - 2.7$. Such steep 
spectra can be obtained for the case of usual injection $q(E) \propto 
(E^{-2} - E^{-2.3})$ from a single source assuming sources 
distribution over maximum acceleration energies $n_s(E_{\max}) 
\propto E_{\max}^{-\beta}$ \cite{KS,Aletal}, where $n_s$ is the 
space density of sources.
\item The generation index $\gamma_g=2.6 - 2.7$ is the main factor which
provides the agreement of proton spectrum with the pair-production dip
and the GZK cutoff (see subsection \ref{dip}). The cosmological evolution
of the sources can be easily included, affecting mostly the low energy
part of the spectrum. Inclusion of additional parameters allows to
improve the fit. In particular, in \cite{BGGprd} the spectrum
was calculated with an account for cosmological evolution of AGN as it 
follows from X-ray observations. The calculated spectrum shown in Fig.~14 
of Ref.~\cite{BGGprd} is in excellent agreement with data.
\item The mode of propagation (from rectilinear to diffusive) is important
in two energy ranges: at low energies $E \lesssim 1$~EeV and at $E
\gtrsim 60$~EeV. In both regimes the distance between sources becomes
important in the diffusive regime. In the low-energy range it makes the
spectrum at $E < E_{\rm tr}$ more flat, ``magnetic flattening'' shown
by curve 'extr.p' in Fig.~\ref{fig:dip-ankle}. And at $E \gtrsim 60$~EeV
the GZK shape becomes more steep in the case of large distance between
sources (see \cite{AB2004} - \cite{BG2007}). 
\end{itemize}

The confirmation of the dip model follows from: ({\em i}) the agreement 
of the dip energy spectrum shape (see Fig.~\ref{fig:dips} and 
Fig.~\ref{fig:PAO-spct}) with observations, ({\em ii}) the agreement of
fluxes of all experiments after the dip-based energy recalibration 
(see Fig.~\ref{fig:PAO-spct} and Refs.\ \cite{Aletal}, Bl\"umer et al.\
from \cite{GZK-rev}, \cite{Yak}), ({\em iii}) the agreement of 
$\left\langle X_{\max}\right\rangle(E)$ with the bulk of observational data in the left panel of
Fig.~\ref{fig:Xmax}, excluding only the highest energy PAO experimental
points. 

A characteristic feature of the dip model is the sharp transition from
Galactic Iron to extragalactic protons. It results in a steep increase
of the $\left\langle X_{\max}\right\rangle(E)$ with energy at $E$ below $1$~EeV 
(see left panels of Fig.~\ref{fig:Xmax}). This feature differentiates the dip model from
the ankle model, where the increase of $\left\langle X_{\max}\right\rangle(E)$ is 
less steep (see the right panel of Fig.~\ref{fig:Xmax}). The difference is due to
less hard proton spectrum at $E \lesssim 1$~EeV in the ankle model, see
Fig.~\ref{fig:dip-ankle}.

The key observation to accept or reject the dip model is the chemical 
composition of UHECR at $E>1$ EeV. In the case of a substantial
admixture of nuclei in the spectrum ($>20\%$) the dip model should be
rejected.

At present the dip model is confirmed by the data of HiRes 
by strong proton-dominance at $E > 1$~EeV, and is in contradiction
with the Auger measurements of $\left\langle X_{\max}\right\rangle(E)$ 
and RMS at $E > 4$~EeV.

\subsection{Mixed composition model}
\label{mixedComposition}

The main concept of the mixed composition model (see Allard et al.\ 
\cite{mixed}, \cite{Allard07}, \cite{Globus07}) is based on the argument
that any acceleration mechanism operating in gas involves different
nuclei in acceleration process and thus the primary flux must have a
mixed composition.

The basic features of the {\em mixed composition model} are as follows
\cite{mixed,Allard07,Globus07}:

\begin{figure*}[t]
\begin{center}
 \begin{minipage}[ht]{63 mm}
 \centering
 \vspace{-2mm}
 \includegraphics[width=69 mm, height=55.5 mm]{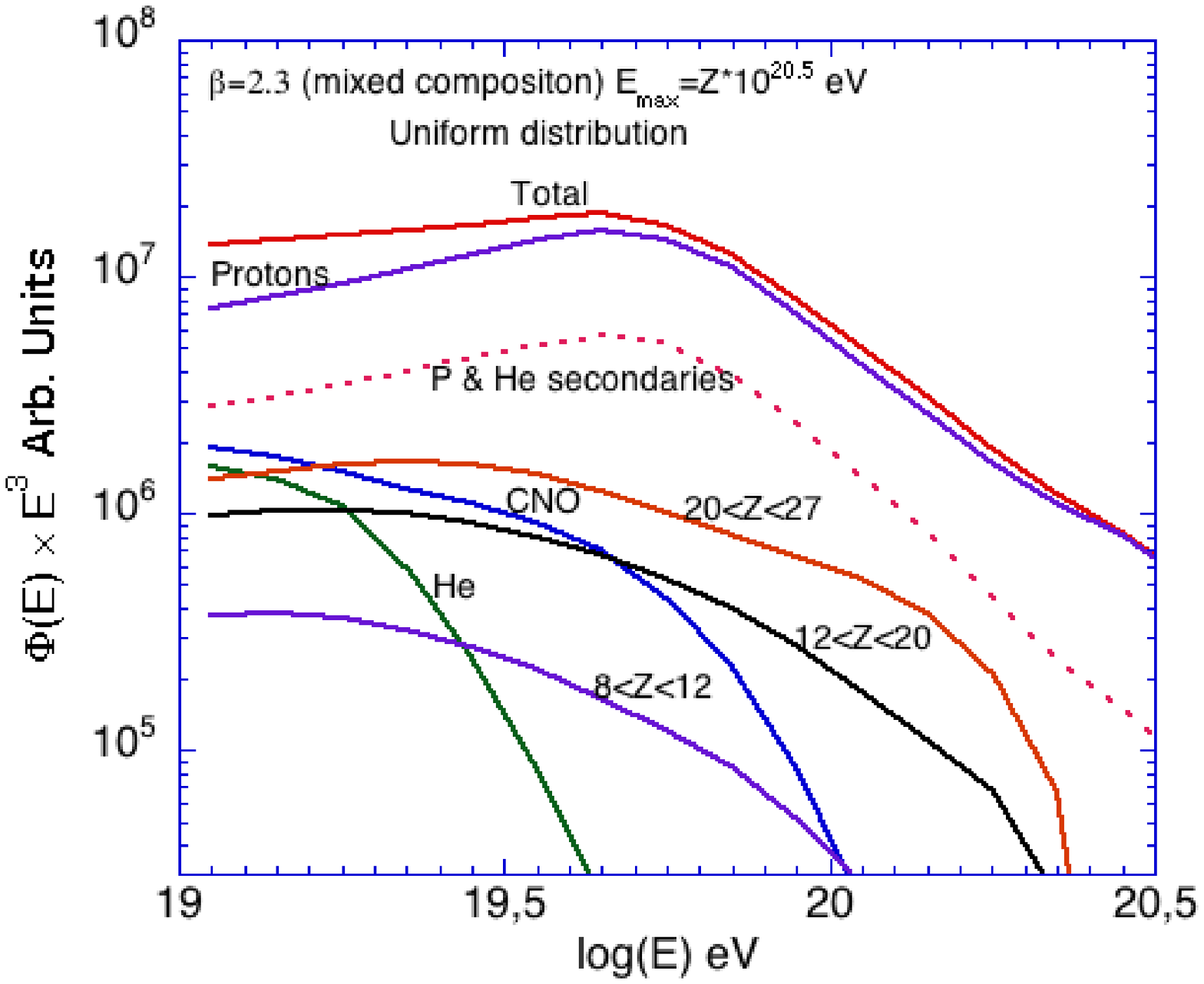}
 \end{minipage}
 \hspace{3mm}
 \begin{minipage}[h]{63 mm}
 \centering
 \includegraphics[width=69 mm, height=54.5 mm]{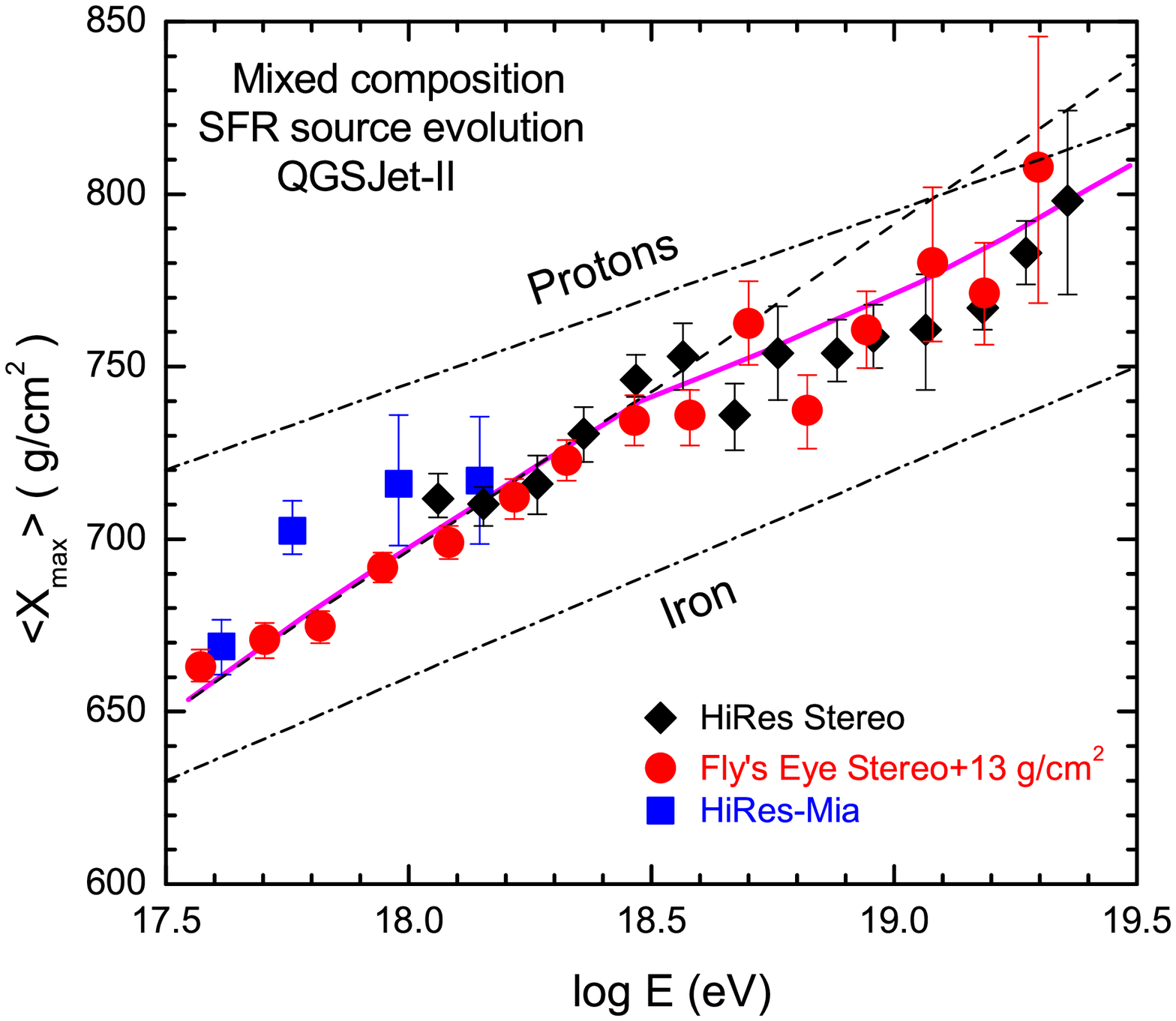}
 \end{minipage}
 \end{center}
\vspace{-4 mm}%
\caption{
{\em Left panel}:~ Diffuse spectrum decomposed in its elemental 
components, calculated in the mixed-composition model by Allard,
Olinto, Parizot (2007) from \cite{mixed}. The proton dominance seen in
the spectrum is a generic prediction of the mixed composition models.
{\em Right panel}:~ Elongation curve $\left\langle X_{\max}\right\rangle(E)$ in the same model as
in the left panel. The mass composition evolves from almost a pure iron
composition at $E \approx 0.3$~EeV to a lighter composition due to an
enrichment by protons and light nuclei of extragalactic origin. At
energy $E_a=3$~EeV the transition to pure extragalactic component is
completed and chemical composition evolution proceeds further due to
photo-disintegration of nuclei. At energy $E \approx 13$~EeV all nuclei
are disappearing faster than before and composition becomes strongly
proton-dominated at $E \geq 30$~EeV.
}
\label{fig:mixed}
\end{figure*} 
\begin{itemize}

\item It is assumed that the Galactic component of CRs above $0.1$ EeV
has a pure iron composition while the extragalactic CRs have a
mixed-nuclei composition. Transition can occur either at ankle or below
it, down to the second knee. Its position is parameter-determined. 

\item In its basic variant \cite{mixed}, the source composition of
extragalactic CRs is assumed to be almost the same as for Galactic CRs,
with protons and Helium being the dominant components. However, a pure 
species composition is also possible in the framework of this model, 
in particular pure Iron, for understanding the physical effects. 

\item The source energy spectra are taken as power-law $Q(E)\propto
E^{-\gamma_g}$ with a generation index $\gamma_g=2.1 - 2.3$. The
maximum acceleration energy is assumed to be rigidity dependent with a
maximum energy for nuclei $i$ being $E^{max}_i =Z_i E^{max}_p$, where
$Z_i$ is the nucleus atomic number. The maximum energy of protons is
typically taken $E^{max}_p = 300$ EeV.

\item The cosmological evolution of sources is described by the factor
$k_{\rm ev}=(1+z)^m$ for a wide range of regimes: $m=0$ in the absence 
of evolution, $m=3$ like for star formation evolution and $m=4$ for a
strong cosmological evolution. These regimes are typically considered
only for small redshifts $z \lesssim 1$ or $1.3$ and for larger $z$ the
frozen evolution with $k_{\rm ev}=\mbox{const}$ up to $z_{\max}$ is
assumed. 

\item The sources are assumed to be distributed homogeneously in the
universe, so that the propagation mode in extragalactic magnetic fields, 
from rectilinear to diffusive, does not influence the energy spectrum at 
observation. The propagation is studied using a Monte Carlo approach with 
the inclusion of pair-production energy losses and nuclei
photo-disintegration on CMB and EBL. 
\end{itemize}

At a first glance one may expect that the large number of free parameters,
such as generation index, parameters of cosmological evolution and
coefficients of source nuclei composition, can provide a broad variety
of observed mass compositions and spectra. However, as it was
demonstrated in \cite{mixed}, the predictions are very much constrained 
due to photo-disintegration of nuclei on EBL and CMB radiations. The 
basic physics phenomena and their results are as follows.

Generically in the mixed models the mass composition becomes lighter at
$E > 10$~EeV, because intermediate and heavy nuclei are destroyed by the EBL
photons while protons survive. In principle this situation may change
only above $50$~EeV, when GZK cutoff in the proton spectrum appears,
while heavy nuclei, e.g.\ Iron, are still not photo-disintegrated by the CMB
photons and may dominate. In realistic cases the dominant component in
mixed models are protons.

In the academic case of a fixed nuclei species at the source (Allard et
al (2008) in \cite{mixed}) secondary protons are the dominant
component at all energies if the primaries are light or intermediate
mass nuclei, and only in the case of Iron the protons are subdominant.
It is interesting to note that because of an efficient destruction of
light and intermediate nuclei the two-component model with only proton
and Iron injected at sources gives a reasonable agreement with
observations \cite{disapp}.

In the mixed models the large primary proton component contributes more
than the secondary protons, strengthening further effect of proton
dominance. Therefore, the GZK feature is also present in the 
mixed-composition models. 

The proton dominance in mixed models is illustrated by 
Fig.~\ref{fig:mixed}. In the left panel decomposition of the diffuse
flux in the elemental components is presented. The dominant component is
primary protons, and the next subdominant is secondary protons and
Helium. In the right panel the elongation curve $\left\langle X_{\max}\right\rangle(E)$ 
is presented with data points from HiRes-MIA, HiRes-stereo and Fly's Eye 
superimposed. Starting from $E \geq 3$~EeV the elongation curve 
$\left\langle X_{\max}\right\rangle(E)$ tends to a proton-dominated composition in 
accordance with HiRes and Fly's Eye data. In contrast, the PAO data tend at the 
highest energies to an Iron-dominated composition. 

\begin{figure}[ht]
\begin{center}
 \includegraphics[width=75 mm]{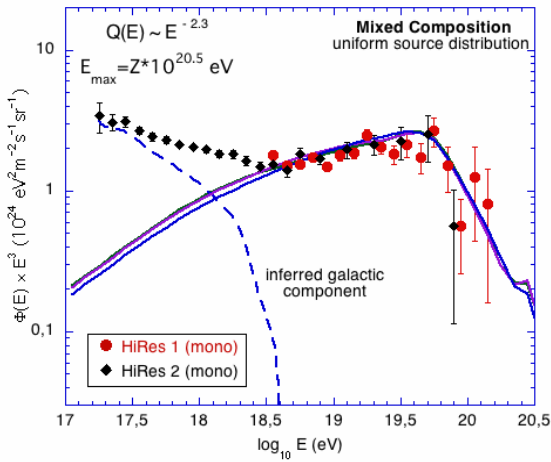}
\end{center}
 \vspace{-1mm}
 \caption{
 Diffuse propagated spectrum for mixed model by Allard et al.\ (2007)
\cite{mixed} with generation index $\gamma_g=2.3$ and cosmological
evolution $(1+z)^3$ up to $z=1.3$ and with a frozen evolution at larger
$z$ up to $z_{\max}=6$. The transition accomplishes at $E \approx
3$~EeV, i.e.\ below the ankle. At highest energies the spectrum is
dominated by protons (see right panel of Fig.~\ref{fig:mixed}) and shows
the GZK cutoff, compared in the figure with HiRes data.
} 
\label{fig:mixed_spctr}
\end{figure}

Transition from Galactic to extragalactic component in the mixed models 
depends on the choice of parameters. In most models transition occurs
at the ankle, see Allard et al.\ in \cite{mixed}. However, in the 
conceptually important paper by Allard, Olinto, Parizot (2007) from 
\cite{mixed} it was emphasized that for strong source evolution and 
flat generation spectra the intersection of Galactic and extragalactic 
components occurs between $0.5$~EeV and $1$~EeV, i.e.\ at the second 
knee, as in the dip model. The transition in this model is shown in 
Fig.~\ref{fig:mixed_spctr}. The transition begins at $2$~EeV and is 
accomplished at $3$~EeV. 

The dominance of protons (as an example see Fig.~\ref{fig:mixed}) was
the reason why the $\left\langle X_{\max}\right\rangle(E)$ predicted by 
mixed composition models is in a better agreement with HiRes data than with 
PAO data. The observations of PAO show that mass composition becomes heavier 
with increasing energy, and thus the existing calculations in the framework 
of mixed models agree better with the HiRes data. 

However, with the recent proposal \cite{parizot}, discussed in
subsection \ref{sec:auger}, the power of the mixed composition model 
for fitting the PAO data may change due to a possible enhancement of 
the heavy nuclei production. 

\section{Discussion and Conclusions} 
\label{conclusions}%
 
The transition from Galactic to extragalactic CRs takes place in the
energy interval $(0.1 - 10)$ EeV. and it gives a key for understanding both
galactic and extragalactic cosmic rays. 

In its low-energy part this range includes the high-energy tail of the
Galactic CRs. The information obtained at these energies on the Galactic
CRs involves the maximum energy of acceleration, the chemical composition,
the anisotropy and the mode of propagation in the Galaxy. It may unveil the 
whole picture of the origin of the Galactic CRs.

The low-energy part of the extragalactic cosmic rays spectrum can
give a key information on the existence of the pair-production dip and
on the propagation of the CRs in extragalactic magnetic fields. 

Therefore, the experimental studies in the transition region $(0.1 - 10)$
EeV are of paramount importance in this field of research, with the
mass composition measured by different methods being probably the most
important task. 

There are four working detectors which cover partially the above-mentioned 
region: KASCADE-Grande \cite{KASCADE-G}, Tunka \cite{tunka},
Yakutsk \cite{Yak} and IceTop/IceCube \cite{IceTop}. There are also
projects to extend the observations of Telescope Array and PAO to low 
energies, $(0.1 - 1)~EeV$ at TALE \cite{TALE} and at LE-Auger 
\cite{LE-Auger}. The Auger detector has a great potential to
explore the low-energy region of the UHECR spectrum. At present there
are already two new detectors at PAO collecting data at this energy; HEAT
\cite{HEAT}, High Elevation Auger Telescope, detecting the fluorescent
light at higher elevation angles; and AMIGA \cite{AMIGA}, Auger Muons
and Infill Ground Array, for the detection of the EAS muon component.
These detectors, together with TALE, Tunka, Yakutsk and IceCube/IceTop
will provide information on all radiations from EAS, including
fluorescent and Cherenkov light, muons and radio radiation.
One may expect that in this way the present  controversy between 
the mass compositions in the HiRes and Auger detectors will be 
unambiguously solved. The recent measurement
\cite{Garcia-GamezICRC,Gazon2012} of the muon-production depth 
$X_{\max}^{\mu}$ and maximum zenith angle $\theta_{\max}$ by 
on-ground Auger detectors is an important step in this direction.

At the low-energy end of the UHECR the {\em energy spectra} are measured
with an unprecedented accuracy for cosmic ray physics. However, in fact
even in experiments with the same technique, like HiRes and Auger, the
{\em energy scales} are different due to systematic errors (see left
panel of Fig.~\ref{fig:PAO-spct}). Such difference is more natural for
different techniques, e.g.\ scintillator and fluorescent ones in TA
detector. The problem of the {\em energy scale} hopefully may be solved
combining all observational methods. However, there is a physical 'standard
candle' for the detector energy calibration, given by the fixed energy
position of the pair-production dip. The recalibration factor $k_{\rm
cal}$ can be found by the spectrum shift to the energy at which the
agreement between observed and predicted dips is the best
\cite{BGGprd,Aletal}. If the observed dip is really caused by
pair-production process, the fluxes measured in all experiments must 
coincide. This is what we see indeed in Fig.~\ref{fig:PAO-spct}, in 
\cite{BGGprd,Aletal,Yak} and in Bl\"umer at al.\ \cite{GZK-rev}.

Since the mass composition is expected to be different for Galactic and
extragalactic CRs, its measurement gives a reliable way to test the
transition. The best method at present to determine the CR chemical
composition is given by measuring $\left\langle X_{\max}\right\rangle(E)$. 
Unfortunately this method suffers from many uncertainties, including 
those in the value of fluorescent yield, absorption of UV light in the atmosphere 
and uncertainties in the interaction models, needed to convert the mass composition 
into the observed $\left\langle X_{\max}\right\rangle(E)$. Systematic errors in
the $\left\langle X_{\max}\right\rangle$ measurement can be as large as 
$20 - 25$~g/cm$^2$, to be compared with the difference of about 100~g/cm$^2$ 
between the $\left\langle X_{\max}\right\rangle(E)$ for protons and Iron. A better 
sensitivity to distinguish different nuclei is given by the width of the 
$\left\langle X_{\max}\right\rangle(E)$ distribution, i.e. \ RMS$(X_{\max})$ \cite{ABBO}. 

The most self-consistent conclusions on the nature of the observed
UHECRs are obtained at present for HiRes detector: it has observed 
both signatures of proton propagating through CMB; namely pair-production 
dip and the GZK cutoff. Moreover, the conclusion about protons as
primaries is confirmed by the direct measurements of proton composition
with the help of  $\left\langle X_{\max}\right\rangle(E)$ and RMS. GZK cutoff is 
also found in the integral spectrum with characteristic energy 
$\log E_{1/2}= 19.73 \pm 0.07$, in excellent agreement with theoretical prediction 
$\log E_{1/2}= 19.72$. Good agreement with the dip energy shape and with $E_{1/2}$ 
allow to conclude that the HiRes basic energy scale is correct. HiRes data
agrees well with the dip model with AGN as the sources and with cosmological evolution 
of AGN taken from X-ray observations. 

The Auger data are quite different. Measurements of mass composition 
using $\left\langle X_{\max}\right\rangle(E)$ and RMS and confirmed by independent data of 
muon-production depth $X_{\max}^{\mu}$ and rise-time of the signal in 
water-Cerenkov tanks indicate nuclei as primaries with steadily 
increasing $\langle A \rangle$ starting from 4~EeV and up to 35~ EeV, where 
mass composition practically reaches pure Iron. The Auger mass composition, 
if confirmed, excludes the dip model and the observed high-energy steepening    
as GZK cutoff. At present there is no nuclei-based model which explains 
simultaneously the Auger energy spectrum $J(E)$ and elongation curve 
$\left\langle X_{\max}\right\rangle(E)$.

There are three models of transition: ankle, dip and mixed composition
one. They differ most notably by the energy of transition and by the
mass composition of the extragalactic component. 

The {\em ankle model} was the first one which appeared in literature. 
It was proposed already in 1960s after the discovery of a flat feature, 
the ankle, at the Volcano Ranch array \cite{linsley}. The transition in 
this model is described in the most natural way, as the intersection of a
steep ($\propto E^{-3.1}$) Galactic heavy-nuclei component (Iron) with a
flat extragalactic proton component with a generation spectrum $\propto
E^{-2}$. The transition, as calculated in different models, occurs
at $E_{\rm tr} \sim (3 - 10)$~EeV. To fill the gap between $0.1$~ EeV 
(Iron knee) and $E_{\rm tr}$ indicated above the ankle models need 
a new yet unknown high energy Galactic component beyond the Standard 
Model for Galactic Cosmic Rays. Recently the ankle model has been 
strongly disfavored by $\left\langle X_{\max}\right\rangle$ measured below the ankle (see right 
panel of the Fig.~\ref{fig:Xmax}) due to small $X_{\max}$ predicted by 
the model. It is also disfavored by the large (not observed) anisotropy 
predicted at $E \sim (1 - 3)$~EeV \cite{anis-ankle}.

The {\em dip model} is based on the pair-production dip induced in the
extragalactic spectrum of protons interacting with the CMB radiation. The
dip is well confirmed by the observations of HiRes, TA, Yakutsk and
AGASA (see Fig.~\ref{fig:dips}). The successful energy calibration of
all observations (see Fig.~\ref{fig:PAO-spct} as an example) using the
fixed energy position of the pair-production dip can be considered as 
another (indirect) evidence in favor of the dip model. 

The transition in the dip model occurs as an intersection of the steep 
Iron Galactic spectrum with a very flat extragalactic spectrum at energy
$(0.1 - 1)$~EeV. The flatness of this spectrum can be provided as a
'magnetic flatness' in the case of diffusive propagation
\cite{Lem2005,AB2005} or due to the sources distribution over
$E_{\max}$ \cite{KS,Aletal}. A robust prediction of the dip model is the
proton-dominated mass composition above $E \gtrsim 1$~EeV. It is
confirmed by the HiRes and TA data; the calculated elongation curve
$\left\langle X_{\max}\right\rangle(E)$ is in good agreement with the bulk 
of data (see left panel of Fig.~\ref{fig:Xmax}). However, it strongly 
contradicts the highest energy Auger data, especially the RMS($X_{\max}$), 
presented in the right panel of Fig.~\ref{fig:massAu}. The Auger - HiRes 
controversy on the mass composition is a crucial point for the survival of 
the dip model. 

In the {\em mixed composition models} the presence of various nuclei is
assumed in the generation spectrum of the extragalactic component. This
assumption is very reasonable, because various nuclei are present in
the gas where acceleration process operates. Mixed composition models 
with a wide choice of parameters are considered in \cite{mixed}. They
include different generation indices, cosmological evolution and most
notably different relative abundances of nuclei at generation. However,
as was explained in subsection \ref{mixedComposition}, a basic feature
common to most mixed composition models is the proton dominance at the
highest energies $E \gtrsim 10$~EeV. This feature is in conflict with
the Auger data at these energies that shows the mass composition getting
progressively heavier with energy increasing and reaching Iron  at
$E \approx 35$~EeV. On the other hand, the mixed composition models fit
reasonably well the HiRes and Fly's Eye data. 
The Galactic spectrum at the highest energies is assumed to be Iron
dominated. The transition energy is a model-dependent quantity, being
determined by $\gamma_g$ and the parameters chosen for sources
cosmological evolution. However, for the flat generation indices
$\gamma_g= 2.1 - 2.3$ and a reasonable choice of the cosmological
evolution, the transition energy is about 3~EeV (as an example see 
Fig.~\ref{fig:mixed_spctr}). The elongation curve shown in the right
panel of Fig.~\ref{fig:mixed} changes the slope at approximately the
same energy, and the mass composition evolves towards the pure proton
composition. The calculated elongation curve in
Fig.~\ref{fig:mixed} agrees well above the transition with the
data of HiRes stereo and Fly's Eye, though nowadays the data of HiRes
agrees better at $E \gtrsim 3$~EeV with a pure proton composition. 

Much better quality of data on the $\left\langle X_{\max}\right\rangle(E)$ 
is needed to distinguish the dip and mixed-composition models by 
$X_{\rm max}$ measurements.

With the recent proposal \cite{parizot} for an enhancement of heavy
nuclei production at the sources, the mixed-composition models may
hopefully describe better the Auger elongation curve 
$\left\langle X_{\max}\right\rangle(E)$ and its RMS.

As short conclusion one may state that while the ankle models of
transition are severely disfavored, mostly by the $\left\langle X_{\max}\right\rangle(E)$ data at 
$E\sim 10^{18}$~eV, the dip and mixed composition models are in a good
agreement with the $\left\langle X_{\max}\right\rangle(E)$ data from 
the Fly's Eye, HiRes and TA detectors. The present problem is the conflict 
of data on mass composition between the HiRes and TA on one side, and the 
PAO on the other. We do expect that problems with mass
composition in the energy range $(0.1 - 3)$~EeV, of fundamental
importance to asses the transition issue, will be soon resolved with
the HEAT, AMIGA, KASCADE-G, TALE, Tunka, Yakutsk and IceCube/IceTop
experiments.

\section*{Acknowledgments}
We are grateful to Johannes Knapp and Sergey Ostapchenko for very useful
discussions and to the Pierre Auger Collaboration for permission to use 
their data \cite{PAO-11,Salamida,PAO-11cc} prior to journal publication.
We thank the anonymous Referee for a careful reading of the paper and 
numerous valuable questions and corrections. The work of V.B. was partly 
supported by the Ministry of Science and Education of Russian Federation 
(agreement 8525).


\begin{thebibliography}{99}

\bibitem{greisen}
K. Greisen, Phys.\ Rev.\ Lett.\ {\bf 16}, 748 (1966).

\bibitem{ZK}
G.~T. Zatsepin, and V.~A. Kuzmin, Pisma Zh.\ Experim.\ Theor.\
Phys.\ {\bf 4}, 114 (1966).

\bibitem{HiRes-GZK}
R. U. Abbasi et al.\ 
Phys.\ Rev.\ Lett.\ {\bf 100}, 101101 (2008);
arXiv:astro-ph/0703099.\\
H.~P.~Dembinski for the Pierre Auger Collaboration, 
arXiv:astro-ph/1107.4809.

\bibitem{TA-GZK}
C.~C.~H.~Jui [Telescope Array Collab.], arXiv:1110.0133;\\
Y.Tsunesada [Telescope Array Collab.], arXiv:1111.2507.

\bibitem{Auger-GZK}
J.~Abraham et al., 
Phys.\ Rev.\ Lett.\ {\bf 101}, 06110.1 (2008).

\bibitem{stecker69}
F.~W.~Stecker, Phys.\ Rev.\ {\bf 180}, 1264 (1969).

\bibitem{BZ71}
V. S. Berezinsky, and G. T. Zatsepin, Sov.\ Journ.\ Nucl.\ Phys.\ 
{\bf 13}, 453 (1971).

\bibitem{BGZ75}
V.~S.~ Berezinsky, S.~I.~Grigorieva, and G.~T.~ Zatsepin, Proc.\ 14th 
ICRC (Munich 1975) {\bf 2}, 711 (1975); %
Astrophys.\ Sp.\ Sci.\ {\bf 36}, 17 (1975).

\bibitem{hillas75}
M.~Hillas, Proc.\ 14th ICRC (Munich 1975) {\bf 2}, 717 (1975).

\bibitem{stecker75}
J.~L.~Puget, and F.~W.~Stecker, Proc.\ 14th ICRC (Munich 1975) 
{\bf 2}, 734 (1975).

\bibitem{GZK-rev}
P.~Bhattacharjee, and G.~Sigl, Phys.\ Rep.\ {\bf 327}, 109 (2000);\\%
M.~Nagano, and A.~A.~Watson, Rev.\ Mod.\ Phys.\ {\bf 72}, 689 (2000);\\%
A.~Letesier-Selvon, and T.~Stanev, Rev.\ Mod.\ Phys.\ {\bf 83}, 907 (2011), 
arXiv: 1103.0031; \\%
K.~Kotera, and A.~V.~Olinto, Ann.\ Rev.\ Astron.\ Astrophys.\ {\bf 49} (2011), 
arXiv:1101.4256; \\%
F. Aharonian, A. Bykov, E. Parizot, V. Ptuskin and A. Watson,
Space Science Reviews {\bf 166} 97 (2012),  arXiv:1105.0131;\\ %
J.~Bl\"umer, R.~Engel, and J.~R.~H\"orandel, Progress in Part.\ and 
Nucl.\ Phys.\ {\bf 63}, 293 (2009);\\%
U. Katz, and Ch.\ Spiering, Prog.Part.Nucl.Phys. {\bf 67} 651 (2012),
arXiv:1111.0507.

\bibitem{BG88}
V.~S. Berezinsky, and S.~I. Grigorieva, Astron.\ Astrophys.\ 
{\bf 199}, 1 (1988).

\bibitem{Stanev2000}
T.~Stanev et al, Phys.\ Rev.\ D {\bf 62}, 093005 (2000).

\bibitem{BGGPL}
V.~Berezinsky, A.~Z. Gazizov, and S.~I. Grigorieva, Phys.\ Lett.\ B
{\bf 612}, 147 (2005); astro-ph/0502550.

\bibitem{BGGprd}
V. Berezinsky, A.~Z. Gazizov, and S.~I.~Grigorieva, Phys.\ Rev.\ D
{\bf 74}, 043005 (2006); hep-ph/0204357.

\bibitem{Aletal}
R. Aloisio, V. Berezinsky, P. Blasi, A. Gazizov, S. Grigorieva, and
B. Hnatyk, Astropart.\ Phys.\ {\bf 27}, 76 (2007); astro-ph/0608219.

\bibitem{data}
R.~U. Abbasi et al.\ [HiRes Collab.], Phys.\ Rev.\ Lett.\
{\bf 92}, 151101 (2004);\\%
V.~P. Egorova et al.\ [Yakutsk Collab.], Nucl.\ Phys.\ B
(Proc.\ Suppl.) {\bf 3}, 136 (2004);\\%
K. Shinozaki et al.\ [AGASA Collab.], Nucl.\ Phys.\ B
(Proc.\ Suppl.) {\bf 3}, 151 (2006); \\%
M. Honda et al.\ [Akeno Collab.], Phys.\ Rev.\ D
{\bf 70}, 525 (1993); \\ %
D.~J. Bird et al.\ [Fly's Eye Collab.], Astrophys.\ J. {\bf 424}, 
491 (1994); \\ %
G.~B.~Thomson [Telescope Array Collab.], arXiv:1010.5528; \\ %
C.~C.~H.~Jui [Telescope Array Collab.], arXiv:1110.0133; \\ %
M.~Roth [Pierre Auger Collab.], astro-ph/0706.2096; \\ %
L.~Perrone [Pierre Auger Collab.], astro-ph/0706.2643. 

\bibitem{HS85}
C.~T. Hill, and D.~N. Schramm, Phys.\ Rev.\ D {\bf 31}, 564
(1985).

\bibitem{YT}
S. Yoshida, and M. Teshima, Progr.\ Theor.\ Phys.\ {\bf 89}, 833
(1993).

\bibitem{linsley}
J.~Linsley, Proc.\ 8th ICRC (Jaipur 1963) {\bf 4} 77 (1963).

\bibitem{auger-ankle}
The Pierre Auger Collaboration, Phys.\ Lett.\ B {\bf 685}, 239 (2010); \\ %
The Pierre Auger Collaboration, ICRC2011, arXiv:1107.4809.

\bibitem{stanev2005}
D. De~Marco, and T. Stanev, Phys.\ Rev.\ D {\bf 72}, 081301
(2005).

\bibitem{waxman}
E. Waxman, Nucl.\ Phys.\ B (Proc.\ Suppl.) {\bf 87}, 345 (2000).

\bibitem{second-knee}
M.~Nagano et al.\ [Akeno Collab.], J.\ Phys.\ G:
Nucl.\ Part.\ Phys.\ {\bf 18}, 423 (1992); \\ %
D.~J.~Bird et al.\ [Fly's Eye Collab.], Phys.\ Rev.\ Lett.\ {\bf71}, 
3401 (1993); \\ %
A.~V.~Glushkov et al.\ [Yakutsk Collab.], JETP Lett.\ {\bf 73}, 115 (2001).

\bibitem{2knee-ankle-rev}
D.~R.~Bergman, and J.~W.~Beltz, J.\ Phys.\ G {\bf 34}, 359 (2007), 
arXiv:0704.3721.

\bibitem{MSU-knee}
G.~Kulikov, and G.~Khristiansen, JETP {\bf 35}, 635 (1958).

\bibitem{kascadeFe}
W.~D.~Apel et al.\ [KASCADE-Grande Collab.], Phys.Rev.Lett. {\bf 107} 171104 (2011), 
arXiv:1107.5885.

\bibitem{mixed}
D. Allard et al., Astron.\ Astrophys.\ {\bf 443}, L29 (2005),
astro-ph/0505566; \\%
D.~ Allard, E.~ Parizot, and A.~ V.~ Olinto, Astropart.\ Phys.\ {\bf 27}, 61 (2007); \\ %
D.~Allard et al.\ J. Phys.\ G {\bf 34}, 359 (2007), astro-ph/0512345; \\ %
D.~Allard et al.\ JCAP 0810:033, (2008), arXiv:0805.4779; \\ %
C.~De Donato, and G.~A.~Medina-Tanko, Astropart.\ Phys.\ {\bf 32}, 253 (2009).

\bibitem{BaaZwi}
W.~Baade, and F.~Zwicky, Proc.\ of National Acad.\ of Scien.\ 
of the United States {\bf 20}, 259, 1934; Phys.\ Rev.\ {\bf 46}, 76 (1934).

\bibitem{GiSy} 
V.~L.~Ginzburg, Usp.\ Fiz.\ Nauk. (Sov.\ Phys.\ Usp.) {\bf 51}, 343 (1953); \\ %
V.~L.~Ginzburg, and Syrovatsky, The Origin of Cosmic Rays, Pergamon
Press, Oxford (1964).

\bibitem{book}
V.~S.~Berezinsky, S.~V.~Bulanov, V.~A.~Dogiel, V.~L.~Ginzburg, and
V.~S.~Ptuskin, Astrophysics of Cosmic Rays, North Holland, Amsterdam, 
534 p.\ (1990).

\bibitem{galprop}
A.~W.~Strong, I.~V.~Moskalenko, and V.~S.~Ptuskin, Ann.\ Rev.\ of
Nucl.\ and Part.\ Science {\bf 57}, 285 (2007).

\bibitem{blasi-rev}
P.~Blasi, arXiv:1012.5005.

\bibitem{DSA}
G.~F.~Krymsky, Sov.\ Phys.\ Dokl.\ Acad.\ Nauk USSR {\bf 243}, 1306 (1977); \\ %
W.~I.~Axford et al.\ Proc.\ 15th ICRC, Plovdiv {\bf 11}, 132 (1977); \\%
A.~R.~Bell, Mon.\ Not.\ RAS {\bf 182}, 147 (1978); \\ %
R.~D.~Blandford, and J.~P.~Ostriker, Astrophys.\ J. {\bf 221}, L229 (1978).

\bibitem{Bell}
A.~R.~Bell, and S.~G.~Lucek, MNRAS {\bf 321}, 433 (2001); \\ %
A.~R.~Bell, MNRAS {\bf 353}, 550 (2004).

\bibitem{Blasi}
E.~Amato, and P.~Blasi, MNRAS {\bf 371}, 1251 (2004); \\ %
E.~Amato, and P.~Blasi, MNRAS Lett.\ {\bf 364}, 76 (2005).

\bibitem{PZ06}
V.~S.~Ptuskin, and V.~N.~Zirakashvili, Advances in Space Research 
{\bf 37}, 1898 (2006).

\bibitem{Ellison}
D.~Ellison, L.~O~.C.~Drury, and J.-P.~Meyer, Astrophys.\ J. {\bf 487}, 197 
(1997).

\bibitem{Meyer}
J.-P.~Meyer, L.~O~.C.~Drury, and D.~Ellison, Astropart.\ Phys.\ J. {\bf 487}, 182
(1997).

\bibitem{BK99}
E.~G.~Berezhko, and L.~T.~Ksenofontov, J. Exp.\ Theor.\ Phys.\ {\bf 89}, 391
(1999).

\bibitem{DBS07}
D.~De Marco, P.~Blasi, and T.~Stanev, JCAP {\bf 0706} 027 (2007), arXiv:0705.1972.

\bibitem{BV07}
E.~G.~Berezhko, and H.~J.~V\"olk, arXiv:0704.1715.

\bibitem{PtuskinDrury}
For a detailed discussion see the papers by V.~Ptuskin, and
L.~O'C.~Drury in this topical issue. 

\bibitem{BlasiAmato}
P.~Blasi, and E.~Amato, JCAP {\bf 1201} 010 (2012), arXiv:1105.4521;\\ 
P.~Blasi, and E.~Amato, JCAP {\bf 1201} 011 (2012), arXiv:1105.4529.

\bibitem{pamela-positrons}
O.~Adriani et al.\ [PAMELA Collab.], Nature {\bf 458}, 607 (2009),
arXiv:08010.4995.

\bibitem{blasi-positrons}
P.~Blasi, Phys.\ Rev.\ Lett.\ {\bf 103}, 051104 (2009).

\bibitem{KOT} 
M.~Kachelrie\ss, S.~Ostapchenko, and R.~Tomas,
Astrophys.\ J. {\bf 733}, 119 (2011).

\bibitem{AharAtoyVolk} 
F. Aharonian, A. Atoyan, and H. J. V\"{o}lk,
 Astron.\ Astroph.\ {\bf 294}, L41 (1995).

\bibitem{MPohl} 
M.~Pohl, Phys.\ Rev.\ D {\bf 79}, 041301 (2009).

\bibitem{e+review}
P.~D.~Serpico, arXiv:1108.4827.

\bibitem{pamela-pHe}
O.~Adriani et al.\ [PAMELA Collab.], Science {\bf 332}, 69 (2011);
arXiv:1103.4055.

\bibitem{drury2010}
L.O'C.~Drury, arXiv:1009.4799.

\bibitem{malkov}
M.~A.~Malkov, P.~H.~Diamond, and R.~Z.~Sagdeev, Phys.Rev.Lett. {\bf 108} 081104 (2012),
arXiv:1110.5335.

\bibitem{ohira}
Y.~Ohira, and K.~Ioka, arXiv:1011.4405.

\bibitem{cesarsky}
P.~O.~Laggage, and C.~J.~Cesarsky, Astron.\ Astrophys.\ {\bf 125}, 249 (1983).

\bibitem{hillas2005}
A.~M.~Hillas, J.\ Phys.\ G. Nucl.\ Part.\ Phys.\ {\bf 31}, R95 (2005).

\bibitem{hillas2006}
A.~M.~Hillas, arXiv:0607109.

\bibitem{gaisser}
T.~Gaisser, Invited talk at XVI Int.\ Symp.\ on Very High
Energy Cosmic Ray Interactions, 2010, Batavia, USA; arXiv:1010.5996.

\bibitem{biermann88}
H.~J.~V\"olk, and P.~L.~Biermann, Astrophys.\ J. {\bf 333}, L65 (1988).

\bibitem{BP89}
V.~S.~Berezinsky, and V.~S.~Ptuskin, Astron.\ Astrophys.\ {\bf 215}, 399 (1989).

\bibitem{sveshnikova}
L.~G.~Sveshnikova, Astron.\ Astrophys.\ {\bf 409}, 799 (2003).

\bibitem{PZS2010}
V.~Ptuskin, V.~Zirakoshvili, and E-S.~Seo, Astrophys.J. {\bf 718} 31 (2010), arXiv: 1006.0034. 

\bibitem{nozawa2010}
T.~Nozawa et al.\ Astrophys.\ J. {\bf 713}, 356 (2010).

\bibitem{dermer}
S.~D.~Wick, C.~D.~Dermer, and A.~Atoyan, Astropart.\ Phys.\ {\bf 21}, 125 (2004).

\bibitem{kusenko}
A.~Calvez, A.~Kusenko, and S.~Nagataki, Phys.\ Rev.\ Lett.\ {\bf 105},
091101 (2010), arXiv:1004.2535.

\bibitem{AloNuclei}
R. Aloisio, V. Berezinsky, and S. Grigorieva, Astrop. Phys. in press 
DOI: 10.1016/j.astropartphys.2012.07.010, arXiv:0802.4452 [astro-ph]; \\ %
R. Aloisio, V. Berezinsky and S. Grigorieva, Astrop. Phys. in press 
DOI: 10.1016/j.astropartphys.2012.06.003, arXiv:1006.2484 [astro-ph.CO].

\bibitem{SteckerEBL} 
F. W. Stecker, M. A. Malkan, and S. T. Scully,
Astrophys.\ J. {\bf 648}, 774 (2006); Astrophys.\ J. {\bf 658}, 1392
(2007);\\ %
T. M. Kneiske, T. Bretz, K. Mannheim, and D.H. Hartmann, Astron.\
Astrophys.\ {\bf 413}, 807 (2004); \\
A.~Franceschini, G. Rodighiero and M. Vaccari, Astron.\ Astrophys.\
{\bf 487}, 837 (2008).

\bibitem{PAO-11} 
P.~Abreu et al.\ [Pierre Auger Collab.], arXiv:1107.4809.

\bibitem{Salamida} 
F. Salamida [Pierre Auger Collab.], 32nd ICRC, Beijing, China, 2011.

\bibitem{Auger-mass}
J.~Abraham et al.\ [Pierre Auger Collab.], Phys.\ Rev.\ Lett.\ {\bf 104},
091101 (2010); \\ %
J.~A.~Bellido et al [Pierre Auger Collab.], Proc.\ 31st ICRC Lodz 2009.

\bibitem{QGSJET}
N.~N.~Kalmykov, and S.~S.~Ostapchenko, A.~I.~Pavlov,
Bull.\ Russ.\ Acad.\ Sci.\ Phys.\ {\bf 58}, 1966 (1994); 
Nucl.\ Phys.\ Proc.\ Suppl.\ {\bf 52B}, 17 (1997).

\bibitem{QGSJETII}
S.~Ostapchenko, Phys.\ Rev.\ D {\bf 83}, 014018 (2011);
Phys.\ Rev.\ D {\bf 74}, 014026 (2006);
Nucl.\ Phys.\ Proc.\ Suppl.\ {\bf 151 B}, 143 (2006).

\bibitem{hires-mia}
R.~U.~Abbasi et al.\ [HiRes-MIA Collab.], Phys. Lett.\ B {\bf 556}, 1 (2003).

\bibitem{E_1/2hires}
R.~U.~Abbasi et al.\ [HiRes Collab.], Phys.Rev.Lett. {\bf 100} 101101 (2008), 
arXiv:astro-ph/0703099.

\bibitem{Yak}
A.~A.~Ivanov, S.~P.~Knurenko, and I.~E.~Sleptsov, New Journ.\ Phys.\ 
{\bf 11}, 065008 (2009).

\bibitem{Garcia-GamezICRC}
D.~Garcia-Gamez, for the Pierre Auger Collaboration, arXiv:astro-ph/1107.4804.

\bibitem{Gazon2012} 
L.~Gazon, for the Pierre Auger Collaboration, J.Phys.Conf.Ser. {\bf 375} 052003 (2012), 
arXiv:astro-ph/1201.6265.

\bibitem{AMIGA}
F.~Sanchez, for the Pierre Auger Collaboration, J.Phys.Conf.Ser. {\bf 375} 052006 (2012),
arXiv:astro-ph/1107.4807.

\bibitem{AllardReview}
D. Allard, arXiv:astro-ph/1111.3290.

\bibitem{Taylor11}
A.M. Taylor, M. Ahlers, F.A. Aharonian, Phys. Rev. {\bf D84} 105007 (2011), 
arXiv:astro-ph/1107.2055.

\bibitem{disapp}
R.~Aloisio, V.~Berezinsky, A.~Gazizov, Astropart.\ Phys.\ {\bf 34}, 620
(2011);\\
J. Phys.\ Conf.\ Ser.\ {\bf 337}, 012042 (2012). 
 
\bibitem{parizot}
C.~Blaksley, E.~Parizot, Astropart.\ Phys.\ {\bf 35}, 342 (2012).

\bibitem{KS}
M. Kachelrie\ss, and D. Semikoz, Phys.\ Lett.\ B {\bf 634}, 143
(2006).

\bibitem{SIBYLL}
R.~S.~Fletcher, T.~K.~Gaisser, P.~Lipari, and T.~Stanev, Phys.\ Rev.\ D
{\bf 50}, 5710 (1994).

\bibitem{PAO-11cc} 
P.~Abreu et al.\ [Pierre Auger Collab.], arXiv:1107.4804.

\bibitem{WW2005}
T.~ Wibig, and A.~W. Wolfendale, J.\ Phys.\ G {\bf 31}, 255
(2005).

\bibitem{BGH}
V.~Berezinsky, S.~Grigorieva, and B.~Hnatyk, Astropart.\ Phys.\ {\bf 22}, 617 (2004).

\bibitem{ABBO}
R.~Aloisio, V.~Berezinsky, P.~Blasi, S.~Ostapchenko, Phys.\ Rev.\ D {\bf 77}, 025007 (2008).

\bibitem{anis-ankle}
G.~Giacinti, M.~Kachelriess, D.~V.~Semikoz, and G.Sigl, JCAP {\bf 1207} 031 (2012),
arXiv:1112.5599.

\bibitem{Lem2005}
M.~Lemoine, Phys.\ Rev.\ D {\bf 71}, 083007 (2005).

\bibitem{AB2005} 
R.~Aloisio, and V.~Berezinsky, Astrophys.\ J. {\bf 625}, 249 (2005).

\bibitem{BGG-AGN}
V.~Berezinsky, A.~Gazizov, and S.~Grigorieva, arXiv:astro-ph/0210095.

\bibitem{berez77}
V.~S.~Berezinsky, Invited lecture at 15th ICRC (Plovdiv, Bulgaria) 
{\bf 10}, 84 (1977).

\bibitem{dermerAGN}
A.~Atoyan, and C.~ D.~ Dermer, New Astron.\ Rev.\ {\bf 48}, 381 (2004), 
astro-ph/0402646.

\bibitem{AB2004}
R.~Aloisio, and V.~Berezinsky, Astrophys.\ J. {\bf 612}, 900 (2004).

\bibitem{BG2007}
V.~Berezinsky, and A.~Gazizov, Astrophys.\ J. {\bf 669}, 684 (2007).

\bibitem{Allard07}
D.~Allard, A.~V.~Olinto, and E,~Parizot, astro-ph/0703633.

\bibitem{Globus07}
N.~Globus, D. Allard, and E.~Parizot, arXiv:0709.1541.

\bibitem{KASCADE-G}
K.-H.~Kampert [KASCADE Grande Collab.], 
Nucl.\ Phys.\ B. (Proc.\ Suppl.) {\bf 122 C}, 422 (2003).

\bibitem{tunka}
B.~V.~Antokhonov et al.\ Texas Symp.\ on Relativistic 
Astroph., Heidelberg, Germany, POS (Texas 2010) 138.

\bibitem{IceTop}
Ice Cube Collab., Proc.\ 32 st ICRC Beijing 2011, 	arXiv:1111.2735v2 [astro-ph.HE].	

\bibitem{TALE}
D.~R.~Bergman [TA/TALE Collab.], Proc.\ of 29th ICRC (Pune)
v.\ 8, 141 (2005); ibid C.~C.~H.~Jui v.\ 8, 101 (2005).

\bibitem{LE-Auger}
M.C. Medina, M. Gomez Berisso, I. Allekotte, A. Etchegoyen,
G. Medina Tanco, A.D. Supanitsky, Nucl.Instrum.Meth. {\bf A566} 302 (2006), 
astro-ph/0607115.

\bibitem{HEAT}
T.~H.~J.~Mathes [Pierre Auger Collab.], J.Phys.Conf.Ser. {\bf 375} 052006 (2012),
arXiv:1107.4807.

\end{thebibliography}
\end{document}